\documentclass[preprint,12pt,authoryear]{elsarticle}%
\usepackage{amssymb,enumitem,booktabs,xcolor,todonotes,microtype,setspace}
\usepackage[top=1.4in, bottom=1.4in, left=1.35in, right=1.35in]{geometry}
\usepackage{natbib}
\usepackage{url}
\usepackage[pdftex,colorlinks=true]{hyperref}
\usepackage{amsmath}
\usepackage{amsfonts}
\usepackage{epsfig}
\usepackage{graphics}
\usepackage{multirow}
\usepackage{graphicx}
\usepackage{amssymb}
\usepackage{palatino,eulervm}
\usepackage{rotating}
\usepackage{dcolumn}
\usepackage{amsthm,thmtools}%
\usepackage{empheq}
\usepackage{comment}
\usepackage{mathabx}
\usepackage{subcaption}

\usepackage{todonotes}

\newcolumntype{L}{D{.}{.}{2,3}}
\definecolor{darkblue}{rgb}{0,0,.6}
\hypersetup{citecolor=darkblue,linkcolor=darkblue,urlcolor=darkblue}
\DeclareMathOperator*{\argmin}{arg\,min}
\DeclareMathOperator*{\argmax}{arg\,max}

\providecommand{\U}[1]{\protect\rule{.1in}{.1in}}

\graphicspath{{plots/}}

\allowdisplaybreaks[4]
\declaretheorem[numberwithin=section]{theorem}
\declaretheorem[numberwithin=section]{lemma}
\makeatletter
\def\th@newremark{\th@remark\thm@headfont{\bfseries}}
\makeatletter
\theoremstyle{newremark}
\newtheorem{remark}{Remark}
\newtheorem{definition}{Definition}
\newtheorem{proposition}[theorem]{Proposition}

\newtheorem{assumption}{Assumption}
\numberwithin{equation}{section}
\declaretheoremstyle[
  spaceabove=6pt, spacebelow=6pt,
  headfont=\bfseries,
  notefont=\mdseries, notebraces={(}{)},
bodyfont=\normalfont,
  postheadspace=0.5em,
  ]{mystyle}

\journal{Journal of Econometrics}
\begin{document}
	\begin{frontmatter}
		\title{Approximate Factor Model with S-vine Copula Structure}
		\author[a]{Jialing Han}
		\ead{jialing.han@york.ac.uk}
		\author[a]{Yu-Ning Li\corref{cor1}}
		\address[a]{School for Business and Society, University of York, UK}
		\cortext[cor1]{Corresponding author.}
		\ead{yuning.li@york.ac.uk}
		
		\begin{abstract}
We propose a novel framework for approximate factor models that integrates an S‐vine copula structure to capture complex dependencies among common factors. Our estimation procedure proceeds in two steps: first, we apply principal component analysis (PCA) to extract the factors; second, we employ maximum likelihood estimation that combines kernel density estimation for the margins with an S‐vine copula to model the dependence structure. Jointly fitting the S‐vine copula with the margins yields an oblique factor rotation without resorting to ad hoc restrictions or traditional projection pursuit methods. Our theoretical contributions include establishing the consistency of the rotation and copula parameter estimators, developing asymptotic theory for the factor-projected empirical process under dependent data, and proving the uniform consistency of the projected entropy estimators. Simulation studies demonstrate convergence with respect to both the dimensionality and the sample size. We further assess model performance through Value-at-Risk (VaR) estimation via Monte Carlo methods and apply our methodology to the daily returns of S\&P 500 Index constituents to forecast the VaR of S\&P 500 index.

		\end{abstract}
		\begin{keyword}
               Distributional forecasting\sep
               Empirical process\sep
		Latent factor\sep
   Principal component analysis\sep
			 Semiparametric copula model

			\emph{JEL Classification}:
			C14\sep
			C32\sep
			C38\sep
			C55.
		\end{keyword}
 	\end{frontmatter}


\newpage


\section{Introduction}\label{sec1}
\subsection{Background}
Latent factor models and copula models are both essential tools in dependence modelling. Latent factor models are widely employed to uncover the underlying structure of high-dimensional data, offering a parsimonious representation of dependencies among large numbers of variables. Their ability to reduce dimensionality has enabled applications across diverse fields, including economics, social sciences, and environmental studies.
In parallel, copula models have emerged as a powerful approach of capturing non-linearity for both serial and cross-sectional dependence among random variables, offering insights that traditional correlation measures may overlook. Vine copulas, in particular, have gained popularity due to their ability to combine multiple bivariate copulas into a cohesive framework, thereby accommodating complex dependence structures. By adding invariance conditions, the stationary vine (S-vine) copula can be obtained, which could facilitate inference in stationary time series.

This paper focuses on the dependence structure of the latent factors of an approximate factor model. Specifically, we consider a high-dimensional random vector, $\mathbf{X}_t$, generated by an approximate factor model:
\begin{equation}\label{eq1.1}
    \mathbf{X}_t = \boldsymbol{\Lambda} \mathbf{F}_t + \boldsymbol{\varepsilon}_t, \quad t = 1, 2, \cdots, n,
\end{equation}
where $\boldsymbol{\Lambda} = (\lambda_1, \cdots, \lambda_d)^\top$ is the matrix of factor loadings, $\mathbf{F}_t$ is an $K$-dimensional vector of common factors, and $\boldsymbol{\varepsilon}_t = (\varepsilon_{t1}, \cdots, \varepsilon_{td})^\top$ represents the idiosyncratic components uncorrelated with $\mathbf{F}_t$. 
By incorporating an S-vine copula structure within the common factors, we aim to extend the model's capacity to capture non-linear and asymmetric dependencies.This allows the model to better accommodate complex interactions between variables that may arise in high-dimensional data. The resulting model is capable of representing more intricate dependence patterns than either factor models or copulas alone.

\subsection{Literature review}

Copula models are widely used to model dependence structures in univariate or multivariate time series, capturing both cross-sectional and serial dependence. As the dimensionality of the time series increases, traditional copula models can become intractably complex. Vine copulas, introduced by \cite{joe1996families}, offer a flexible framework by decomposing a multivariate copula into a sequence of bivariate copulas structured over a series of nested trees. The development of vine copula models has been advanced in a number of works, including \cite{cooke1997markov}, \cite{joe2014dependence}, \cite{kurowicka2006uncertainty}, \cite{ACFB09},\cite{loaiza2018time} and \cite{zhao2022modeling}.


Various types of vine structure exist, such as D-vines, C-vines, and more general R-vines. In the context of time series modelling, \cite{CF06a} investigate first-order Markov models as a special case of D-vines. \cite{ibragimov2009copula} extends D-vine constructions to higher-order Markov models. For multivariate higher-order Markov chains, several alternative structures have been proposed, including D-vines \citep{smith2015copula}, M-vines \citep{beare2015vine}, and COPAR models \citep{brechmann2015copar}. More recently, \cite{NKM22} establish general conditions under which a vine copula model yields a stationary multivariate time series. They show that D-vines and M-vines satisfy a translation invariance condition that ensures stationarity. While COPAR structures may be stationary in some cases, they do not satisfy this condition in general.


To address the challenge of modelling dependence in high-dimensional data, copula models have been extended to incorporate factor structures. \cite{joe2010tail} introduces the factor copula model, in which the observed variables are assumed to be conditionally independent given a latent factor. This framework is generalized by \cite{krupskii2013factor} to allow for multiple latent factors.  
A Gaussian factor model implies a Gaussian
factor copula, but factor copulas allow for non-Gaussian dependence, enabling richer modelling of tail behaviour and asymmetry. 
These models have been widely applied in finance and insurance, as discussed in \cite{embrechts2009copulas},  \cite{krupskii2013factor}, \cite{chen2015multi} and others.  A special case is the linear factor copula model studied by \cite{OP17} and \cite{KG18}, where observations are linear in factors and idiosyncratic errors. To handle dynamic dependence, \cite{OP18} propose copula-based time-varying factor models, and \cite{opschoor2021closed} extend this to dynamic factor copulas with observation-driven time-varying loadings.

In parallel, the statistical literature on latent factor models offers a complementary approach to modeling high-dimensional data through dimension reduction. While factor copula models typically rely on conditional independence given latent variables, classical factor models assume that the errors are uncorrelated with the factors and with one another. A key advancement is the approximate factor model introduced by \cite{CR83}, which allows for weakly correlated idiosyncratic errors. When both the number of variables and the sample size are large, approximate factor models can be efficiently estimated using principal component analysis (PCA) \citep{BN02}.
This approach is extensively employed in economics \citep[e.g.,][]{SW02} and enhances large covariance matrix estimation \citep[e.g.,][]{FLM13}, which is crucial for effective risk management in finance. 

Our approach adopts the approximate factor model framework but shifts focus toward understanding the dependence structure of the common latent factors. In contrast to factor copula models, which impose a conditional independence assumption between observed variables given the latent factors, the approximate factor model places less assumptions on the dependence structure between factors and idiosyncratic errors. Moreover, while the approximate factor model is primarily designed to capture conditional mean relationships, our work instead emphasizes the full distributional properties of the latent factors, including both their cross-sectional and serial dependence. This perspective allows us to capture temporal and cross-sectional dynamics in the underlying factors  beyond what traditional mean-based models reveal.

The semiparametric estimation of copula models with non-parametric margins and parametric copula functions is studied by \cite{CF06a,CF06b}, where the marginal distributions and the copula can be estimated separately. In our setting, however, the marginal distributions and the copula share a common set of parameters, so a full‐likelihood approach is required.
Empirical process theory plays a central role in developing the asymptotic theory of maximum likelihood estimation for our model. 
\cite{KWXXY19} study the empirical process in approximate factor models under an i.i.d. factor assumption, whereas our work extends these results to accommodate dependent factors under a mixing condition. 
To resolve the factor rotation indeterminacy, we examine the projected empirical process of factors, a tool often used in model diagnostics \citep[e.g.,][]{ZL98, E06, X09}. The consistency of our MLE hinges on the uniform convergence of this projected empirical process, specifically of the projections $\boldsymbol{v}^\top \boldsymbol{F}_t$, over all unit vectors $\boldsymbol{v}$. 


Our estimation of the marginal log-likelihood is closely related to entropy estimation; see \cite{MP24} for a comprehensive review. Entropy estimators are widely used in projection pursuit to identify informative directions in high-dimensional data \citep[e.g.,][]{Huber85}. In particular, for a one-dimensional projection, one may search for a direction $\boldsymbol{v}$ that maximizes the non-normality by maximizing the entropy $I(\boldsymbol{v}) = -\int g_{\boldsymbol{v}} \log g_{\boldsymbol{v}}$, where $g_{\boldsymbol{v}}$ denotes
the density of the scalar projection $\boldsymbol{v}^\top \boldsymbol{F}_t$. 
\cite{Joe89} proposes the integral-form plug‐in entropy estimators based on kernel density estimation, with the leave-one-out variant discussed in \cite{IR81}.  \cite{HM93} further extends both estimators without resorting to numerical integration. 
 In particular, \cite{HJWW20} establish minimax rates for entropy estimation and demonstrate that integral-form kernel-based plug-in estimators are nearly optimal, exhibiting only logarithmic suboptimality.  However, most of the literature assumes i.i.d. samples and only a few studies consider dependent data \citep[e.g.,][]{Lim07,RAMM15}.  In this paper, we use entropy estimators to estimate the log-likelihood of the marginal distributions of rotated latent factors. 
Although the uniform consistency of the entropy estimator $I(\boldsymbol{v})$ over $\boldsymbol{v}$ has been established under i.i.d.\ assumptions \citep[e.g.,][Theorem 1]{ZWF95} and more recently in kernel-based likelihood estimation for independent component analysis by \cite{HHW25}, we extend these results by establishing the uniform consistency of kernel-based plug-in entropy estimators under dependence and by accounting for the estimation error introduced through principal component analysis when estimating latent factors.

\subsection{Contribution}

In this paper, we propose a new semiparametric framework that integrates the approximate factor model with an S-vine copula structure to investigate the dependence among latent common factors. Our method extends principal component analysis by introducing a rotation of the initial factors that aligns them with a semiparametric vine copula model in the second step. This additional step enables the model to accommodate complex, non-Gaussian dependence structures that standard PCA fails to capture. 

Unlike traditional approaches that rely on projection pursuit techniques such as quartimax or oblimax \citep[see][]{Huber85}, or on imposed structural constraints like sparsity in factor loadings \citep[e.g.,][]{DD23}, our procedure performs a likelihood-based estimation that simultaneously determines an optimal oblique rotation and fits a flexible S-vine copula model.
As a result, our method captures rich dependence patterns across time and variables and facilitates distributional forecasting.

Our theoretical contributions are threefold. First, we establish consistency for both the factor rotation and the copula parameter estimates. Second, we develop the asymptotic theory for the empirical process of the latent factors in the presence of temporal dependence, extending existing results to mixing-dependent sequences. Third, we derive uniform convergence results for kernel-based entropy estimators, accounting for the compounded effects of projection, factor estimation error, and serial dependence. These developments collectively support the validity and robustness of our proposed estimation framework.

\subsection{Organization of the Paper}
The remainder of the paper is organized as follows. Section~\ref{sec3} introduces the approximate factor model with S-vine copula-dependent common factors and outlines a two-step estimation procedure. Then we explores distributional forecasting based on the estimated model. Section~\ref{sec4} establishes the asymptotic properties of the estimators. In particular, we study the convergence of the empirical process of the estimated factors under arbitrary rotations and prove the consistency of the proposed estimators. Section~\ref{sec5} presents numerical results and Section~\ref{sec6} illustrates an empirical application using real data. Section~\ref{sec7} concludes.

\ref{appendixA} contains the proofs of the main asymptotic theorems, while \ref{appendixB} provides additional technical lemmas and their proofs. \ref{appendixC} discusses the parameterization of the rotation matrix.

\medskip

Throughout the paper, we use $\Vert\cdot\Vert$ to denote the Euclidean norm of a vector, the operator (or spectral) norm of a matrix, or the supremum norm of a function, depending on the context.  Let $\lambda_{\max}(\cdot)$, $\lambda_{\min}(\cdot)$, $\operatorname{Tr}(\cdot)$, and $\operatorname{det}$ denote the maximum eigenvalue, minimum eigenvalue, trace, and determinant of a square matrix, respectively. For a $p\times p$ matrix ${\mathbf W}=(w_{ij})_{p\times p}$, we let $\Vert {\mathbf W}\Vert_F=\operatorname{Tr}^{1/2}\left({\mathbf W}^{^\intercal}{\mathbf W}\right)$ be its Frobenius norm, $\vert{\mathbf W}\vert_1=\sum_{i=1}^p\sum_{j=1}^p |w_{ij}|$, $\Vert{\mathbf W}\Vert_1=\max_{1\leq j\leq p}\sum_{i=1}^p |w_{ij}|$, $\Vert{\mathbf W}\Vert_{\infty}=\max_{1\leq i\leq p}\sum_{j=1}^p |w_{ij}|$, and $\Vert {\mathbf W}\Vert_{\max}=\max_{1\leq i\leq p}\max_{1\leq j\leq p} |w_{ij}|$.  Denote $\nabla_{\boldsymbol{x}}{f}$ as the gradient of the function $f$ with respect to the vector $\boldsymbol{x}$.
For a generic vector of functions \( \boldsymbol{G} = (G_1, \cdots, G_K) \), write $\boldsymbol{G}(\boldsymbol{x}) = (G_1(x_1), \cdots, G_K(x_K))$ with $\boldsymbol{x}=(x_1,\cdots,x_K)$.

\section{Model and Estimation Procedure}\label{sec3}

In this section, we first introduce a factor model with S-vine copula dependence common factors and then we develop a two-step estimation procedure.
Suppose we have time series observations $X_{it}$ for $t = 1, \cdots , T$ and $i = 1,\cdots, N$. Our goal is to estimate the copula structure of the latent factors. 

\subsection{Model setting: Factor model with S-vine copula dependence common factors}


To accommodate strong cross-sectional dependence which is not
uncommon for large-scale time series collected in practice, we assume that $X_{it}$ is generated by an
approximate factor model:
\begin{equation}\label{eq3.1}
X_{it}= \boldsymbol{\lambda}_i^\top{\boldsymbol F}_t+\varepsilon_{it},
\end{equation}
where $\boldsymbol{\lambda}_i$ is a $K$-dimensional vector of the factor loading, ${\boldsymbol F}_t=(F_{1t},\cdots,F_{Kt})^\top$ is a $K$-dimensional vector of latent
factors, and $\varepsilon_{it}$ is an idiosyncratic error.
In matrix notation, \eqref{eq3.1} can be written as
\begin{equation}\label{eq3.2}
{\bf X}=  {\bf F} \boldsymbol{\Lambda}^\top + \boldsymbol{\varepsilon},
\end{equation}
where ${\bf X}$ is a $T$-by-$N$ matrix, ${\bf F}  = ({\boldsymbol F}_1,\cdots, {\boldsymbol F}_T)^\top$ is a $T$-by-$K$ matrix, $\boldsymbol{\Lambda} = (\boldsymbol{\lambda}_1,\cdots, \boldsymbol{\lambda}_N)^\top$ is a $N$-by-$K$ matrix , and $\boldsymbol{\varepsilon}$ is a $T$-by-$N$ error matrix. 

In order to capture the dependence structure among factors, we assume that the factors of the approximate factor model follow a Markovian S-vine copula model \citep{NKM22}.

\begin{definition} \label{as1}
$({\boldsymbol F}_1,\cdots, {\boldsymbol F}_T)$ follows a \emph{Markovian S-vine copula model} $({\cal V}^F,{\cal C}({\cal V}^F))$ of order $p$, where
 ${\cal V}^F$ is the vine structure consisting of a collection of trees $T_k$ with vertex sets $V_k^F$ and edge sets $E_k^F$, and
 ${\cal C}({\cal V}^F)$ is the collection of pair copulas associated with the edges in the vine, if the following conditions are satisfied:

(i)  The edges of the $(k-1)$-th tree will be the vertices of the $k$-th tree, where $V_k^F=E_{k-1}^F$ for $k=2,\cdots, Kp$.

(ii) Assuming vertices $a,b \in V_k^F$ are connected by an edge $e\in E_k^F$ and the corresponding edges $a=\{a_1,a_2\},b=\{b_1,b_2\}\in E_{k-1}^F$, there is exactly one of the element in $a$ equals one of the element in $b$. 

(iii) There are $Kp-1$ conditioning set for for any edges in the $(Kp-1)$-th tree $T_{Kp-1}$. There is no conditioning set in the first tree. Every conditioned set appears exactly once.

(iv)  Assuming vertices $a_e,b_e \in V_k$ are connected by an edge $e\in E_k^F$, and edges $a_{e'}$ and $b_{e'}$ are the corresponding edges shift in time by $\tau$ steps, that is, $a_e = a_{e'} + (0, \tau )$ and $b_e = b_{e'} + (0, \tau )$, the pair copulas is translation invariant, that is, $c_{a_e, b_e | {\cal D}_e} = c_{a_{e'}, b_{e'} | {\cal D}_{e'}}$.

\medskip

Here, we adopt standard notation from the vine copula literature. For any edge \( e \in E_k \), the complete union of \( e \), denoted \( U_e \), is defined as
\[
{\cal U}_e = \left\{ i \in V_1 \,\middle|\, i \in e_1 \in e_2 \in \cdots \in e \text{ for some } (e_1, \ldots, e_{k-1}) \in E_1 \times \cdots \times E_{k-1} \right\}.
\]
For a singleton \( i \in V_1 \), we define \( {\cal U}_i := \{i\} \). For an edge \( e \in E_k \) connecting two nodes \( v_1 \) and \( v_2 \), we define the \emph{conditioning set} as \(  {\cal D}_e := {\cal U}_{v_1} \cap {\cal U}_{v_2} \), and the \emph{conditioned set} as the pair \( (a_e, b_e) := ({\cal U}_{v_1} \setminus  {\cal D}_e, {\cal U}_{v_2} \setminus {\cal D}_e) \). We label such an edge by \( e = (a_e, b_e \mid  {\cal D}_e) \).

\cite{NKM22} investigate different S-vine structures, including M-vines, which are proposed by \cite{beare2015vine} as a special case of S-vine. Within an M-vine framework, the cross-sectional dependence at each time point is represented by a D-vine configuration, wherein the D-vines corresponding to periods $t$ and $t+1$ are interconnected along a common boundary of their respective vine trees.
Figure \ref{Fig:2} shows a three dimensional M-vine on three time points, where each column is a D-vine that captures the cross-sectional dependence.

\begin{figure}[h!]
    \centering
    \includegraphics[width = 0.4\textwidth]{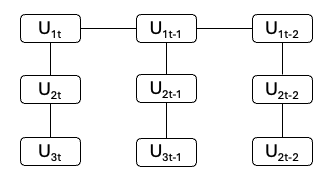}
    \caption{Example of first tree level of three-dimensional M-vine on three time points, where $U_{it}$ denotes the probability integral transform of $F_{it}$.}
    \label{Fig:2}
\end{figure}

\end{definition}
\medskip

\subsection{Estimation procedure: PCA and full maximum likelihood estimation}\label{sec3.2}
 To estimate the copula structure of the common factors, we first estimate the unobserved factors ${\bf F}$ using principal component analysis. Specifically,
we estimate the unobserved factors using $\widehat{\bf F}= (\widehat{{\boldsymbol F}}_1,\cdots, \widehat{{\boldsymbol F}}_T)^\top$ which is
the matrix of $K$ eigenvectors (multiplied by $\sqrt{T}$) associated with the $K$ largest eigenvalues of the matrix ${\bf X}{\bf X}^\top/(TN)$ in decreasing order. Without loss of generality, we assume that the first non-zero element of each eigenvector is positive. The factor loading matrix is estimated by $\widehat{\boldsymbol{\Lambda}}={\bf X}^\top\widehat{\bf F}/T$. In practice, the number of factors, $K$, is usually unknown, and we determine it via an information criterion \citep[]{BN02},  
which minimizes the following function over different values of $k$,
\begin{equation*}
	\operatorname{IC}(k)=\log \left(\sum_{t=1}^T\Vert{\boldsymbol{X}}_t-\widehat{\bf F}(k) \widehat{\boldsymbol{\Lambda}}^\top(k)\Vert^2\right)+k\cdot\left(\frac{T+N}{TN}\right)\log (T\wedge N).
\end{equation*}

Due to the identification issue inherent in factor models, the decomposition ${\bf X} = {\bf F} \Lambda + \boldsymbol{ \varepsilon}$ is not unique. Specifically, for any invertible matrix ${\bf H}$, the model can equivalently be written as ${\bf X} = ({\bf F}{\bf H})({\bf H}^{-1}\Lambda) + \boldsymbol{ \varepsilon}$. As a result, the estimated factors $\widehat{\bf F}$ approximate ${\bf F}$ only up to a rotation. Therefore, directly fitting $\widehat{\bf F}$ to an S-vine copula model may not yield a consistent estimate of the true vine structure.

To address this, we parametrize the rotation matrix as ${\bf H} = {\bf H}({\boldsymbol \theta}_H)$ and jointly estimate both the rotation parameters ${\boldsymbol \theta}_H$ and the copula parameters ${\boldsymbol \theta}_C$ via the maximum likelihood estimation. We assume that the likelihood function admits a unique global maximum (see Assumption \ref{as4}). 
Specifically, the expected (qausi-)log-likelihood function of ${\bf F}_1,\cdots,{\bf F}_T$,  
$$ T\log(\operatorname{det}({\bf H}({\boldsymbol \theta}_H)))+T L_{M,HF}({\boldsymbol \theta}_H)+L_{C,HF}({\boldsymbol \theta}_C,{\boldsymbol \theta}_H)$$ 
attains a unique maximum at $({\boldsymbol \theta}_H^\circ, {\boldsymbol \theta}_C^*)$ with ${\bf H}({\boldsymbol \theta}_H^\circ)={\bf I}_K$, where  $L_{M,HF}$ and $L_{C,HF}$ denotes the expected marginal log-likelihood and the expected copula log-likelihood function, respectively, based on the rotated factors ${\bf F}{\bf H}({\boldsymbol \theta}_H)$.

Noting that ${\bf H}$ is the Jacobian matrix \citep[e.g.,][pp.53--55]{Wi62} of the variable transformation from ${\boldsymbol F}_t$ to ${\bf H}{\boldsymbol F}_t$, ${\bf H}({\boldsymbol \theta}_H^\circ)={\bf I}_K$ implies that the log-likelihood function is uniquely maximized when no rotation is applied to the true factors. This ensures the identifiability of the true rotation and guarantees consistency in jointly estimating the rotation and copula parameters. Since the parameterization of the rotation matrix is not unique, the uniqueness of the maximum depends on the specific parameterization. We demonstrate that this identifiability is achievable through simulation in Section~\ref{sec4.1}. In practice, we normalize each column of the rotation matrix to have unit norm and carefully handle the potential sign-flip issue especially when the copula family is asymmetric. Further discussion on the parameterization of the rotation matrix is provided in \ref{appendixC}.


Assume that the PCA estimator $\widehat{\bf F}$ is an consistent estimator of ${\bf F}{\bf R}_0$, where ${\bf R}_0$ is a deterministic rotation matrix which is defined later in Proposition \ref{prop1}. Define the c.d.f. and p.d.f. of ${\boldsymbol H}_j^\top{\bf R}_0{\boldsymbol F}_t$ as $G_j(x;{\boldsymbol \theta}_H)$ and $g_{jt}(x;{\boldsymbol \theta}_H)$, respectively, for $j=1,\cdots,K$, where ${\boldsymbol H}_j$ is the $j$-th column of ${\bf H}$.
 We can define the empirical distribution function of ${\boldsymbol H}_j^\top{\bf R}_0{\boldsymbol F}_t$ as follows,
\begin{equation}\label{eq3.3}
\widehat G_j(x;{\boldsymbol \theta}_H)= \frac{1}{T+1}\sum_{t=1}^T {\boldsymbol 1}({\boldsymbol H}_j^\top\widehat{\boldsymbol F}_t\leq x),
\end{equation}
for $j=1,\cdots,K$,
and the pseudo-observations as 
$$\widehat {\boldsymbol U}_{t}({\boldsymbol \theta}_H)=\widehat {\boldsymbol G}({\bf H}\widehat{\boldsymbol F}_t; {\boldsymbol \theta}_H)=(\widehat  G_1({\boldsymbol H}_j^\top\widehat{\boldsymbol F}_t; {\boldsymbol \theta}_H),\cdots,\widehat  G_K({\boldsymbol H}_j^\top\widehat{\boldsymbol F}_t; {\boldsymbol \theta}_H))^\top.$$ 
To implement maximum likelihood estimation, we need also estimate the negative differential entropy of ${\boldsymbol H}_j^\top{\bf R}_0{\boldsymbol F}_t$, that is $\operatorname{E}[\log(g_{j}({\boldsymbol H}_j^\top{\bf R}_0{\boldsymbol F}_t;{\boldsymbol \theta}_H))]$. 
This can be estimated by using the kernel density estimator \citep[e.g.,][]{HMSW04},
$$\widehat g_{j}(x;{\boldsymbol \theta}_H)= \frac{1}{Tb_j}\sum_{s=1}^T W(({\boldsymbol H}_j^\top\widehat{\boldsymbol F}_s-x)/b_j),$$
or the leave-one out kernel density estimator,
$$\widehat g_{jt}(x;{\boldsymbol \theta}_H)= \frac{1}{(T-1)b_j}\sum_{s\neq t} W(({\boldsymbol H}_j^\top\widehat{\boldsymbol F}_s-x)/b_j),$$
which results in the leave-one out kernel entropy estimator \citep[e.g.,][]{MBND24},
\begin{equation}
\widehat L_{G_j}({\boldsymbol \theta}_H):=\frac{1}{T}\sum_{t=1}^T\log\left(\widehat g_{jt}({\boldsymbol H}_j^\top\widehat{\boldsymbol F}_t;{\boldsymbol \theta}_H)\right),
\end{equation}
%
for $j=1,\cdots,K$, where $W$ is a kernel function and $b_j$ is the bandwidth. 
\cite{Hall88} argues that for the nonnegative kernels we should use a window of size between $T^{- 1/3}$ and $T^{- 1/4}$ for entropy estimation. Thus we choose $b_j=\widehat\sigma_j T^{-1/4}$ as the bandwidth, where $\widehat\sigma_j$ is the standard deviation estimate of ${\boldsymbol H}_j^\top\widehat{\boldsymbol F}_t$. 
Then, the likelihood function of ${\bf F}$ can be defined as
\begin{equation}\label{eq.LLH}
\widehat L_{T,{\bf b}}(\widehat{\bf F}{\bf H};{\boldsymbol \theta}_H,{\boldsymbol \theta}_C)=\log(|{\bf H}|)+\widehat L_{M}({\boldsymbol \theta}_H)+\frac{1}{T}\widehat L_C(\widehat {\boldsymbol U}_1({\boldsymbol \theta}_H),\cdots, \widehat {\boldsymbol U}_T({\boldsymbol \theta}_H);{\boldsymbol \theta}_C),
\end{equation}
with bandwidth ${\boldsymbol b}=(b_1,\cdots,b_K)$ and  $\widehat L_{M}({\boldsymbol \theta}_H)=\sum_{j=1}^K\widehat L_{G_j}({\boldsymbol \theta}_H)$ being the estimate of $L_{M}({\boldsymbol \theta}_H):=\sum_{j=1}^K\operatorname{E}[\log(g_{j}({\boldsymbol H}_j^\top{\bf R}_0{\boldsymbol F}_t;{\boldsymbol \theta}_H))]$ and $\widehat L_C$ is the log-likelihood of the S-vine copula which is defined later in \eqref{eq2.7}.

By maximizing the objective function in \eqref{eq.LLH}, we obtain the estimates $\widehat{\boldsymbol \theta}_C$ and $\widehat{\boldsymbol \theta}_H$. Let $\widehat{\bf H} = {\bf H}(\widehat{\boldsymbol \theta}_H)$. The factors are then recovered as $\widehat{\bf F}\widehat{\bf H}$ with the S-vine copula parameters $\widehat{\boldsymbol \theta}_C$.
In practice, the vine structure is usually unknown, and we follow the procedure as in Appendix S3.1 of \cite{NKM22} to determine it, also see \cite{DBCK13}.

\subsection{Summary and further discussion}
We summarize the estimation procedure:
\begin{description}
    \item[Step 1:] Obtain $\widehat{\boldsymbol F}_t$, for $t=1,\cdots,T$ using PCA as discussed at the beginning of Section \ref{sec3.2}.
    \item[Step 2:]  For given 
${\boldsymbol \theta}_H$, calculate the pseudo-observations, $\widehat {\boldsymbol U}_{t}({\boldsymbol \theta}_H)$ and estimate the parameter $\widehat{\boldsymbol \theta}_C({\boldsymbol \theta}_H)$ for the S-vine copula.
 Then the estimator of ${\boldsymbol \theta}_H$ is defined as 
\begin{equation}
\widehat{\boldsymbol \theta}_H=\argmax_{{\boldsymbol \theta}_H \in{\boldsymbol \Theta}_H} \widehat L_{T,{\bf b}}(\widehat{\bf F}{\bf H};{\boldsymbol \theta}_H,\widehat{\boldsymbol \theta}_C({\boldsymbol \theta}_H)),
\end{equation}
where $\widehat L_{T,{\bf b}}$ is defined in \eqref{eq.LLH}. Then we let $\widehat{\boldsymbol \theta}_C=\widehat{\boldsymbol \theta}_C(\widehat{\boldsymbol \theta}_H)$. Thus, we obtain the estimates of the original factors and factor loadings (with S-vine dependence structure), respectively, 
$$\widetilde{\bf F}=\widehat{\bf F}{\bf H}(\widehat{\boldsymbol \theta}_H)\ \ \textit{ and }\ \ \widetilde{\boldsymbol{\Lambda}}:=\widehat{\boldsymbol{\Lambda}}({\bf H}(\widehat{\boldsymbol \theta}_H)^{-1})^\top.$$
\end{description}

\medskip

Before concluding this section, we discuss the relationship between the stepwise estimation and Z-estimation. We say that two edges \( e \) and \( e' \) are \emph{translation invariance} of each other if  
\[
a_e = a_{e'} + (\tau, 0), \quad b_e = b_{e'} + (\tau, 0), \quad  {\cal D}_e = \{ v + (\tau, 0) : v \in  {\cal D}_{e'} \}.
\] 
We write \( e \sim e' \) to indicate that \( e \) and \( e' \) are translation invariance. This defines an equivalence relation on the set of edges. The corresponding \emph{equivalence class} of an edge \( e \) is denoted by
\[
[e] := \{ e' : e' \sim e \}.
\]

Since joint maximization of the full likelihood is often computationally intensive, following \cite{NKM22}, we adopt the stepwise maximum likelihood estimator of \cite{ACFB09}, which estimates the parameters of each pair-copula sequentially. Accordingly, we define
\begin{equation}\label{eq2.7}
\widehat L_C(\boldsymbol{\theta}_C, \boldsymbol{\theta}_H) = \sum_{[e']} \widehat L_{C,[e']}(\boldsymbol{\theta}_{[e']}, \boldsymbol{\theta}_H|{\boldsymbol{\theta}}_{S_a([e'])},{\boldsymbol{\theta}}_{S_b([e'])}),
\end{equation}
and for each representative edge \( e' \in E_k^F \), we define
\begin{eqnarray*}
&&\widehat L_{C,[e']}(\boldsymbol{\theta}_{[e']}, \boldsymbol{\theta}_H|{\boldsymbol{\theta}}_{S_a([e'])},{\boldsymbol{\theta}}_{S_b([e'])}) \nonumber\\
&=&
\sum_{e \sim e'} \log c_{[e]} \left\{
C_{a_{[e]} \mid D_{[e]}}\left(\widehat{U}_{a_e} \mid \widehat{\boldsymbol U}_{D_e}; {\boldsymbol{\theta}}_{S_a([e])} \right),
C_{b_{[e]} \mid D_{[e]}}\left(\widehat{U}_{b_e} \mid \widehat{\boldsymbol U}_{D_e}; {\boldsymbol{\theta}}_{S_b([e])} \right); \boldsymbol{\theta}_{[e']}
\right\},
\end{eqnarray*}
where \( \boldsymbol{\theta}_C \) denotes the collection of all \( \boldsymbol{\theta}_{[e]} \), and ${\boldsymbol{\theta}}_{S_a([e])}$ and ${\boldsymbol{\theta}}_{S_b([e])}$ denote the parameter estimates from previous step.

\medskip

Define
\begin{equation*}
\begin{aligned}
& C_{[e], 1, {\boldsymbol \theta}_C}\left(\boldsymbol{u}_t, \cdots, \boldsymbol{u}_{t+p}\right)=C_{a_{[e]} \mid D_{[e]}}\left(u_{a_e} \mid \boldsymbol{u}_{D_{[e]}} ; \boldsymbol{\theta}_{S_a([e])}\right), \\
& C_{[e], 2, {\boldsymbol \theta}_C}\left(\boldsymbol{u}_t, \cdots, \boldsymbol{u}_{t+p}\right)=C_{b_{[e]} \mid D_{[e]}}\left(u_{a_e} \mid \boldsymbol{u}_{D_{[e]}} ; \boldsymbol{\theta}_{S_b([e])}\right),
\end{aligned}
\end{equation*}
and
\begin{equation*}
\boldsymbol{s}_{[e],{\boldsymbol \theta}_C}\left(\boldsymbol{u}_t, \cdots, \boldsymbol{u}_{t+p}\right)=\nabla_{{\boldsymbol \theta}_{[e]}}\log(c_{[e]}(C_{[e], 1, {\boldsymbol \theta}_C}\left(\boldsymbol{u}_t, \cdots, \boldsymbol{u}_{t+p}\right),C_{[e], 2, {\boldsymbol \theta}_C}\left(\boldsymbol{u}_t, \cdots, \boldsymbol{u}_{t+p}\right))).
\end{equation*}
Let ${\boldsymbol s}_{{\boldsymbol \theta}_C, \boldsymbol {\nu}}$ be a vector of functions stacking $\boldsymbol{s}_{[e],{\boldsymbol \theta}_C}$ for $[e]\in E_i$ and $i=1,\cdots, (p+1)K$ and denote ${\boldsymbol s}_{{\boldsymbol \theta}_C, \boldsymbol {\nu}}\left(\boldsymbol{x}_t, \cdots, \boldsymbol{x}_{t+p}\right)={\boldsymbol s}_{{\boldsymbol \theta}_C}\left(\boldsymbol {\nu}(\boldsymbol{x}_t), \cdots, \boldsymbol {\nu}(\boldsymbol{x}_{t+p})\right)$. For given ${\boldsymbol \theta}_H$, we can rewrite the semiparametric step-wise MLE as the Z-estimation as \cite{NKM22}, that is, $\widehat{\boldsymbol \theta}_C({\boldsymbol \theta}_H)$ solves the following estimation equation,
\begin{equation*}
\frac{1}{T-p}\sum_{t=1}^{T-p}{\boldsymbol s}_{{\boldsymbol \theta}_C, \widehat{\boldsymbol G}}(\widehat{\boldsymbol F}_t,\cdots,\widehat {\boldsymbol F}_{t+p}) = {\boldsymbol  0}.
 \end{equation*}

Furthermore, let
\begin{equation*}
\begin{aligned}
&\boldsymbol{s}_{1,{\boldsymbol \theta}_H}=\nabla_{{\boldsymbol \theta}_H}\log(\operatorname{det}({\bf H}({\boldsymbol \theta}_H))),\\
&\boldsymbol{s}_{2,{\boldsymbol \theta}_H}=\nabla_{{\boldsymbol \theta}_H}L_{M}({\boldsymbol \theta}_H),\\
&\widehat{\boldsymbol{s}}_{2,{\boldsymbol \theta}_H}=\nabla_{{\boldsymbol \theta}_H}\widehat L_{M}({\boldsymbol \theta}_H).
\end{aligned}
\end{equation*} 
and ${\boldsymbol s}_{{\boldsymbol \theta}_H, \boldsymbol {\nu}}$ be a vector stacking $\boldsymbol{s}_{[e],{\boldsymbol \theta}_H,{\boldsymbol \nu}}$ for $[e]\in E_i$ and $i=1,\cdots, (p+1)K$, where
\begin{eqnarray}
&&\boldsymbol{s}_{[e],{\boldsymbol \theta}_H,{\boldsymbol \nu}}\left(\boldsymbol{x}_t, \cdots, \boldsymbol{x}_{t+p}\right)\notag\\
&& =\nabla_{{\boldsymbol \theta}_{H}}\log(c_{[e]}(C_{[e], 1, {\boldsymbol \theta}_C}\left(\boldsymbol {\nu}(\boldsymbol{x}_t), \cdots, \boldsymbol {\nu}(\boldsymbol{x}_{t+p})\right),C_{[e], 2, {\boldsymbol \theta}_C}\left(\boldsymbol {\nu}(\boldsymbol{x}_t), \cdots, \boldsymbol {\nu}(\boldsymbol{x}_{t+p})\right))).\notag
\end{eqnarray}
Then our proposed estimator can represent a Z-estimator of the following estimation equation,
\begin{empheq}[left=\empheqlbrace]{align}
&\frac{1}{T-p}\sum_{t=1}^{T-p}{\boldsymbol s}_{{\boldsymbol \theta}_C,\widehat{\boldsymbol{G}}}(\widehat{\boldsymbol F}_t,\cdots,\widehat {\boldsymbol F}_{t+p}) = {\boldsymbol  0}, \nonumber \\
&\boldsymbol{s}_{1,{\boldsymbol \theta}_H}+\widehat{\boldsymbol{s}}_{2,{\boldsymbol \theta}_H}+\frac{1}{T-p}\sum_{t=1}^{T-p}  \boldsymbol{s}_{{\boldsymbol \theta}_H,\widehat{\boldsymbol{G}}}\left(\widehat{\boldsymbol F}_t, \cdots, \widehat{\boldsymbol F}_{t+p}\right)={\boldsymbol  0}. \nonumber
\end{empheq}

\subsection{Forecasting}\label{sec2.4}

Following \cite{NKM22}, we construct the predictive distribution via Monte Carlo simulation. Recall that $\widetilde{\boldsymbol{F}}_t$ and $\widetilde{\boldsymbol{\Lambda}}$ denote the estimated factors and the factor loadings matrix after proper rotation, respectively, and that $\widehat{\boldsymbol{\varepsilon}}_t$ denotes the residuals at time $t$. Our forecasting procedure is as follows:

\begin{description}
    \item[Step 1:] Construct the empirical marginal distribution of $\widetilde{\boldsymbol{F}}_t$, and calculate the rank of  $\widetilde{\boldsymbol{F}}_t$ for $t=1,\cdots, T$, and denoted as  $\widetilde{\boldsymbol{U}}_t$.
    
    \item[Step 2:] Simulate $M$ paths $(\widetilde{\boldsymbol{U}}_{T+1}^{(m)}, \cdots,\widetilde{\boldsymbol{U}}_{T+h}^{(m)})$ for $m=1,\cdots, M$,
 from the S-vine copula model with starting values $\widetilde{\boldsymbol{U}}_{1}, \cdots,\widetilde{\boldsymbol{U}}_{T}$, where $h$ is the forecast horizon. 
        
     \item[Step 3:]  Transform $(\widetilde{\boldsymbol{U}}_{T+1}^{(m)}, \cdots,\widetilde{\boldsymbol{U}}_{T+h}^{(m)})$  into $(\widetilde{\boldsymbol{F}}_{T+1}^{(m)},\cdots, \widetilde{\boldsymbol{F}}_{T+h}^{(m)})$, for $m=1,\cdots, M$, in factor space by quantile mapping via the empirical marginal distribution of $\widetilde{\boldsymbol{F}}_t$.

    \item[Step 4:] Simulate $\widetilde{\boldsymbol{X}}_{T+k}^{(m)} = \widetilde{\boldsymbol{\Lambda}} \widetilde{\boldsymbol{F}}_{T+k}^{(m)} + \widetilde{\boldsymbol{\varepsilon}}_{T+k}^{(m)}$ for $k=1,\cdots,h$ and $m=1,\cdots,M$.
    where $\widetilde{\boldsymbol{\varepsilon}}_{T+k}^{(m)}$ is resampled from $\{\widehat{\boldsymbol{\epsilon}}_1,...,\widehat{\boldsymbol{\epsilon}}_T\}$ with replacement.

\item[Step 5:] 
The predictive distribution at time $T+k$ is then given by the ensemble $\widetilde{\boldsymbol{X}}_{T+k}^{(m)}$ for $m=1,\cdots, M$. Quantiles of this distribution provide prediction intervals, while the ensemble mean serves as the point forecast.
\end{description}

Alternative approaches for forecasting the idiosyncratic errors $\boldsymbol{\varepsilon}_t$ can also be considered. For instance, one may fit univariate time series models, such as ARMA-GARCH with specified innovation distributions, to each component of $\boldsymbol{\varepsilon}_t$. While such methods capture serial dependence within each series, they neglect potential cross-sectional dependence across components.

\section{Asympotics}\label{sec4}

We next list some regularity conditions which are sufficient for deriving the uniform consistency results for $\widehat{\boldsymbol F}_t$, $\widehat{\boldsymbol\lambda}_i$, and $\widehat\varepsilon_{t,i}$.

\begin{assumption}\label{as2}
{ (i) The process $\{({\boldsymbol F}_{t}^\top, {\boldsymbol\varepsilon}_t^\top)^\top \}_{t=1}^\infty$ is stationary and $\beta$-mixing with the mixing coefficient $\beta_l$ satisfying $\beta_l=O\left(\rho^l\right)$, where $\rho$ is a constant that satisfies $ 0<\rho<1$.}

{ (ii) There exists a positive definite matrix ${\boldsymbol\Sigma}_\Lambda$ such that
\[
\left\Vert \frac{1}{N}\sum_{i=1}^N {\boldsymbol\lambda}_{i}{\boldsymbol\lambda}_{i}^\top-{\boldsymbol\Sigma}_\Lambda\right\Vert=O\left(N^{-1/2}\right).
\]
and $\Vert{\boldsymbol\lambda}_{i}\Vert$ is uniformly bounded over $1\leq i\leq N$, where ${\boldsymbol\lambda}_{i}$ is the $i$-th column of ${\boldsymbol\Lambda}^\top$. }

{ (iii) The covariance matrix ${\boldsymbol\Sigma}_F:=\mathsf{Var}({\boldsymbol F}_t)$ is positive definite. In addition, there exists a constant $c_1>0$ such that $\operatorname{E}\left[\exp\left(c_1\Vert {\boldsymbol F}_t\Vert^2\right)\right]<\infty$.}

{ (iv) The idiosyncratic errors satisfy 
$$\operatorname{E}[{\boldsymbol\varepsilon}_t]={\bf 0},\ \ \ \operatorname{E}[\varepsilon_{t,i}{\boldsymbol F}_t]={\bf 0},\ \  \ {\rm and}\   \max_{1\leq i\leq N}\max_{1\leq t\leq T}\operatorname{E}\left[\exp\left(c_1 \varepsilon_{t,i}^2\right)\right]<\infty,$$ where $c_1$ is defined in (iii).}

{ (v) There exist $0<c_2<\infty$ and $\delta_{\varepsilon}>2$ such that}
\[
\max_{1\leq t\leq T}\operatorname{E}\left[\left\Vert\sum_{i=1}^N {\boldsymbol\lambda}_{i}\varepsilon_{t,i}\right\Vert^{\delta_{\varepsilon}}\right]\leq c_2 N^{\delta_{\varepsilon}/2},\ \ \max_{1\leq s,t\leq T}\operatorname{E}\left[\left|\sum_{i=1}^N \left[\varepsilon_{s,i}\varepsilon_{t,i}-\operatorname{E}\left(\varepsilon_{s,i}\varepsilon_{t,i}\right)\right]\right|^{\delta_{\varepsilon}}\right]\leq c_2 N^{\delta_{\varepsilon}/2}.
\]

(vi) {\em The eigenvalues of ${\boldsymbol\Sigma}_\Lambda^{1/2}{\boldsymbol\Sigma}_F{\boldsymbol\Sigma}_\Lambda^{1/2}$ are bounded and distinct.}

(vii) {\em 
For any unit vector $h \in \mathbb{R}^K$, the density function of the projection $h^\top {\boldsymbol F}_t$ exists and is denoted by $g(x)$. 
This density has bounded support $\operatorname{supp} g$, satisfies the following conditions for some $\delta > 0$:
\[ \sup_{\| h \| = 1}\sup_{x\in \operatorname{supp}g} g(x)\geq c >0, \ \ \ \ \sup_{\| h \| = 1} \sup_{x \in \operatorname{supp}g} \left| \nabla_x g(x) \right| < \infty\]
and
\[
\sup_{\| h \| = 1} \operatorname{E} \left[ \left| \log\left( g(h^\top {\boldsymbol F}_t) \right) \right|^{1+\delta} \right] < \infty
.
\]
}
\end{assumption}

\medskip

Most of the above assumptions are standard in the approximate factor model estimation theory \citep[e.g.,][]{BN02, FLM13, CLLL25}. The sub-Gaussian moment conditions in Assumption \ref{as2}(iii) and (iv) are required to cover the ultra-high dimensional case where $N$ diverges at an exponential rate of $T$, and can be weakened if $N$ diverges at a polynomial rate of $T$. Assumption \ref{as2}(vi) ensures the identification of the eigenvectors and Assumption \ref{as2}(vii) are required to estimate the differential entropy of the rotated factors \citep[e.g.,][]{HHW25}.

\smallskip

Define a $K\times K$ (stochastic) rotation matrix 
\begin{equation}\label{eq4.1}
{\bf R}_{NT}:={\boldsymbol V}_{NT}^{-1}\left(\frac{1}{T}\sum_{t=1}^T\widehat{\boldsymbol F}_t{\boldsymbol F}_t^\top\right)\left(\frac{1}{N}\sum_{i=1}^N{\boldsymbol\lambda}_i{\boldsymbol\lambda}_i^\top\right),
\end{equation}
where ${\boldsymbol V}_{NT}$ is a $K\times K$ diagonal matrix with the $K$ largest eigenvalues of ${\mathbf X}{\mathbf X}^\top/(NT)$ (arranged in a descending order) on its main diagonal. The following proposition gives the uniform consistency results for $\widehat{\boldsymbol F}_t$, $\widehat{\boldsymbol\lambda}_i$, and $\widehat\varepsilon_{t,i}$, which are comparable to those obtained in the existing literature \citep[e.g.,][]{BN02, FLM13,LLF23,CLLL25}.
\smallskip

\begin{proposition}\label{prop1}
{ Suppose that Assumption \ref{as2}(i)--(v) is satisfied, $T^{4/\delta_{\varepsilon}} \ll N \ll \exp\{T^{1/5}\}$ with $\delta_{\varepsilon}$ defined in Assumption \ref{as2}(v). Then, we have the following uniform consistency results:
\begin{equation}
\max_{1\leq t\leq T}\left\Vert\widehat{\bf{F}}_t-{\bf R}_{NT}{\bf{F}}_t\right\Vert=O_P\left(\frac{(\log T)^{1/2}}{T}+\frac{T^{2/\delta_{\varepsilon}}}{N^{1/2}}\right).\label{eq4.2}
\end{equation}
Furthermore, if Assumption \ref{as2}(vi) is satisfied, we have
\begin{equation}\label{eq4.3} 
\left\Vert{\bf R}_{NT}-{\bf R}_0\right\Vert =O_P\left(\frac{1}{T^{1/2}}+\frac{T^{2/\delta_{\varepsilon}}}{N^{1/2}}\right), 
\end{equation}
and, as a consequence, 
\begin{equation}\label{eq4.4}
\max_{1\leq t\leq T}\left\Vert\widehat{\bf{F}}_t-{\bf R}_0{\bf{F}}_t\right\Vert=O_P\left((\log T)^{1/2}\left(\frac{1}{T^{1/2}}+\frac{T^{2/\delta_{\varepsilon}}}{N^{1/2}}\right)\right).
\end{equation}
where ${\bf R}_0={\boldsymbol V}_0^{-1/2}{\bf U}_0^\top{\boldsymbol \Sigma}_\Lambda^{1/2}$,
     ${\bf U}_0$ is a matrix consisting of the eigenvectors of ${\boldsymbol\Sigma}_\Lambda^{1/2}{\boldsymbol\Sigma}_F{\boldsymbol\Sigma}_\Lambda^{1/2}$, 
    $v_{i,0}$ is the $i$-th largest eigenvalue of ${\boldsymbol\Sigma}_\Lambda^{1/2}{\boldsymbol\Sigma}_F{\boldsymbol\Sigma}_\Lambda^{1/2}$, and
  ${\boldsymbol U}_0=\operatorname{diag}\{v_{1,0},\cdots,v_{q,0}\}$. 
    }
\end{proposition}

\medskip

Proposition \ref{prop1} shows that $\widehat{\boldsymbol F}_t$ are consistent estimators of the rotated latent factors ${\bf R}_{0}{\bf{F}}_t$, rather than the factors themselves (unless ${\bf R}_{0}=\bf{I}$). In addition, if we assume that $N\gg T^{4/\delta_{\varepsilon}+1}$, the term $T^{2/\delta_{\varepsilon}}/N^{1/2}$ in (\ref{eq4.2}) would disappear. 

Recall that $\widehat G_j(x;{\boldsymbol \theta}_H)=T^{-1}\sum_{t=1}^T {\boldsymbol 1}({\boldsymbol H}_j^\top\widehat{\boldsymbol F}_t\leq x)$ is the marked empirical process of the common factors. 
Define $\widetilde G_j(x;{\boldsymbol \theta}_H)= T^{-1}\sum_{t=1}^T {\boldsymbol 1}({\boldsymbol H}_j^\top{\bf R}_0{\boldsymbol F}_t\leq x)$ as the oracle empirical process of the common factors.
We have the following theorem.

\begin{theorem}\label{thm4.1}
Suppose that Assumptions \ref{as2} are satisfied and $T^{4/\delta_{\varepsilon}+1} \ll N \ll \exp\{T^{1/5}\}$ with $\delta_{\varepsilon}$ defined in Assumption \ref{as2}(v). We have 
\begin{equation}\label{eq4.5}
\sqrt{T}\sup_{{\boldsymbol \theta}_H \in {\boldsymbol \Theta}_H}\sup _{x \in \mathbb{R}}\left|\widehat G_{j}(x;{\boldsymbol \theta}_H)-
\widetilde G_j(x;{\boldsymbol \theta}_H)+\frac{1}{T}\sum_{t=1}^Tg_{j}(x;{\boldsymbol \theta}_H)z_{{\boldsymbol \theta}_H t}\right| \rightarrow_P 0.
\end{equation}
where $z_{{\boldsymbol \theta}_H t}={\boldsymbol H}_j^\top(
\widehat{\boldsymbol F}_t-{\bf R}_0{\boldsymbol F}_t).$ In particular, we have 
\begin{equation}\label{eq4.6}
\sup_{{\boldsymbol \theta}_H \in {\boldsymbol \Theta}_H}\sup _{x \in \mathbb{R}}\left|\widehat G_{j}(x;{\boldsymbol \theta}_H)-
G_j(x;{\boldsymbol \theta}_H)\right|\rightarrow_P 0.
\end{equation}
\end{theorem}

\begin{remark}
 \citet{KWXXY19} establish the consistency of the marked empirical processes for the common components and idiosyncratic errors in an approximate factor model estimated via principal component analysis, under the assumption that the factors are independent and identically distributed (i.i.d.) random vectors, independent of the idiosyncratic component series. In comparison, our paper provides a novel consistency result for the marked empirical processes based on arbitrary rotations of the estimators of the common factors, relaxing these assumptions to accommodate a mixing condition.
\end{remark}
\medskip

Some further assumptions are needed for the maximum likelihood estimation.

\begin{assumption} \label{as3}
(i) The kernel $W(\cdot)$ is a twice continuously differentiable, symmetric probability density function with compact support on the interval $[-1, 1]$.\\
(ii) The bandwidth $b$  satisfies
$b \to 0$ and  $\ Tb^2/ \log^2(T) \to \infty$.
\end{assumption}
\medskip

Define the pseudo-true values ${\boldsymbol \theta}^*=(({\boldsymbol \theta}^*_H)^\top,({\boldsymbol \theta}^*_C)^\top)^\top$ as the solution to 

\begin{equation*}
\operatorname{E}\left[{\boldsymbol \phi}_{\boldsymbol \theta,\boldsymbol {G}}({\bf R}_0{\boldsymbol F}_1,\cdots,{\bf R}_0{\boldsymbol F}_{1+p})+{\boldsymbol \psi}_{\boldsymbol \theta}\right]=\boldsymbol 0,
\end{equation*}
where 
\begin{equation}\label{eq4.7}
{\boldsymbol \phi}_{\boldsymbol \theta,\boldsymbol {\nu}}({\boldsymbol x}_1,\cdots,{\boldsymbol x}_{1+p})=\left(
\begin{aligned}
&{\boldsymbol s}_{{\boldsymbol \theta}_C,\boldsymbol {\nu}}({\boldsymbol x}_t,\cdots,{\boldsymbol x}_{t+p})  \\
&\boldsymbol{s}_{{\boldsymbol \theta}_H,\boldsymbol {\nu}}\left({\boldsymbol x}_t, \cdots, {\boldsymbol x}_{t+p}\right)
\end{aligned}\right)
\end{equation}
and 
\begin{equation}
{\boldsymbol \psi}_{{\boldsymbol \theta}_H}=\nabla_{\boldsymbol \theta}\log(\operatorname{det}({\bf H}({\boldsymbol \theta}_H)))+\nabla_{\boldsymbol \theta}L_{M}({\boldsymbol \theta}_H)=
\left(
\begin{aligned}
&\boldsymbol 0 \\
&\boldsymbol{s}_{1,{\boldsymbol \theta}_H}+ \boldsymbol{s}_{2,{\boldsymbol \theta}_H}
\end{aligned}\right).
\end{equation}
\smallskip
\begin{assumption} \label{as4}
(i) The pseudo-true values $(({\boldsymbol \theta}^*_H)^\top,({\boldsymbol \theta}^*_C)^\top)^\top$ 
lie in the interior of $\Theta_H\times \Theta_C$ and for every $\epsilon>0$,
$$\inf_{\Vert {\boldsymbol\theta}-{\boldsymbol\theta}^*\Vert>\epsilon}\Vert \operatorname{E}\left[{\boldsymbol \phi}_{\boldsymbol \theta,\boldsymbol {G}}({\bf R}_0{\boldsymbol F}_1,\cdots,{\bf R}_0{\boldsymbol F}_{1+p})+{\boldsymbol \psi}_{{\boldsymbol \theta}_H}\right]\Vert>0.$$

(ii) The function ${\boldsymbol \phi}_{\boldsymbol{\theta},\boldsymbol{G}}$
is continuously differentiable with respect to $\boldsymbol\theta$ and its arguments and there is $\delta>0$ such that
\begin{equation*}\operatorname{E}\left[\sup_{{\boldsymbol \theta}\in {\boldsymbol \Theta}} \left\Vert {\boldsymbol \phi}_{\boldsymbol{\theta},\boldsymbol{G}}({\bf R}_0{\boldsymbol F}_1,\cdots,{\bf R}_0{\boldsymbol F}_{1+p})\right\Vert<\infty,\quad \sup_{{\boldsymbol \theta}\in {\boldsymbol \Theta}}\left\Vert \nabla_{\boldsymbol{\theta}}{\boldsymbol \phi}_{\boldsymbol{\theta},\boldsymbol{G}}({\bf R}_0{\boldsymbol F}_1,\cdots,{\bf R}_0{\boldsymbol F}_{1+p})\right\Vert\right]<\infty,
\end{equation*}
and
\begin{equation*}\operatorname{E}\left[\sup_{{\boldsymbol \theta}\in {\boldsymbol \Theta}}\sup_{{\boldsymbol{\nu}}\in \mathcal{F}_\delta} \left\Vert \frac{\partial}{\partial \{\nu_{j}({\boldsymbol \theta}_1,{\bf R}_0{\boldsymbol F}_s)\}}{\boldsymbol \phi}_{\boldsymbol{\theta},\boldsymbol{\nu}}({\bf R}_0{\boldsymbol F}_1,\cdots,{\bf R}_0{\boldsymbol F}_{1+p})\right\Vert\right]<\infty,
\end{equation*}
for $j=1,\cdots,K$, and $s=1,\cdots,1+p$, where
$$\mathcal{F}_\delta=\{{\boldsymbol{\nu}}:\Vert{\boldsymbol{\nu}}-{\boldsymbol{G}}\Vert\leq\delta\}.$$

(iii) The matrix derivatives $\nabla_{\boldsymbol{\theta}}\nabla_{\boldsymbol{x}^\dag}{\boldsymbol \phi}_{\boldsymbol{\theta},\boldsymbol{G}}({\boldsymbol x}^\dag)$
 are continuous in ${\boldsymbol x}^\dag$ where ${\boldsymbol x}^\dag=({\boldsymbol x}_1,\cdots,{\boldsymbol x}_{1+p})$. There is $\delta>0$ such that 
  \begin{equation*}
  \sup_{{\boldsymbol{\nu}}\in \mathcal{F}_\delta} \Vert\nabla_{\boldsymbol{x}^\dag}{\boldsymbol \phi}_{\boldsymbol{\theta},\boldsymbol{\nu}}\Vert<\infty
  \end{equation*}
 and
 \begin{equation*}\operatorname{E}\left[\sup_{{\boldsymbol \theta}\in {\boldsymbol \Theta}}\sup_{{\boldsymbol{\nu}}\in \mathcal{F}_\delta} \left\Vert \frac{\partial}{\partial \{\nu_{j}({\boldsymbol \theta}_H,{\bf R}_0{\boldsymbol F}_s)\}}\nabla_{\boldsymbol{\theta}}{\boldsymbol \phi}_{\boldsymbol{\theta},\boldsymbol{\nu}}({\bf R}_0{\boldsymbol F}_1,\cdots,{\bf R}_0{\boldsymbol F}_{1+p})\right\Vert\right]<\infty,
\end{equation*}
for $j=1,\cdots,K$, and $s=1,\cdots,1+p$.



\end{assumption}

Assumption \ref{as4}(i) ensures the identifiability of the model parameters and \ref{as4}(iv) is a standard regularity
condition. Assumption \ref{as4}(ii)(iii) are standard in the semiparametric copulas model estimation theory \citep[e.g.,][and references therein]{NKM22}.

\begin{theorem}\label{thm4.2}
Suppose that Assumptions \ref{as1}--\ref{as4} are satisfied and $T^{4/\delta_{\varepsilon}+1} \ll N \ll \exp\{T^{1/5}\}$ with $\delta_{\varepsilon}$ defined in Assumption \ref{as2}(v). 
We have 
$$\widehat{\boldsymbol \theta}_H {\rightarrow_p} {\boldsymbol \theta}^*_H,\ \
\widehat{\boldsymbol \theta}_C{\rightarrow_p} {\boldsymbol \theta}^*_C,$$
and 
\begin{equation}
\max_{1\leq t\leq T}\left\Vert{\bf H}(\widehat{\boldsymbol \theta}_H)\widehat{\boldsymbol{F}}_t-{\boldsymbol{F}}_t\right\Vert=O_P\left((\log T)^{1/2}\left[\left(\frac{\log N}{T}\right)^{1/2}+\frac{T^{2/\delta_{\varepsilon}}}{N^{1/2}}\right]\right).
\end{equation}
\end{theorem}

\section{Simulation}\label{sec5}

In this section, we first discuss the practice issue in identification and then present two simulated examples to examine the finite-sample performance of our proposed methods. For each example, we assess the accuracy of the estimate when the number of common factors and the structure of the S-vine are correctly specified.

\subsection{Simulation study on identification issues}\label{sec4.1}
 The S-vine structure and the copula families remain invariant under scaling, permutation, and sign-flipping transformations (see Proposition \ref{prop0} in \ref{appendixC}). While this invariance holds theoretically, practical implementations necessitate restricting the search to a finite set of copula families. We validate this proposition through simulation studies and explore whether additional transformations could introduce identification challenges when the set of admissible copula families is restricted.

\begin{figure}[h!]
    \centering
    \includegraphics[width = 0.25\textwidth]{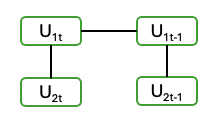}
    \caption{Vine structure of first tree of first-order two-dimensional M-vine, where $U_{it}$ denotes the probability integral transform of $F_{it}$.}
    \label{fig:vs1}
\end{figure}

To illustrate, we simulate two factors generated from an S-vine copula with 1,000 observations. The vine structure is demonstrated in Figure \ref{fig:vs1}. The copula families considered include Gaussian, Clayton, Frank, and Joe, while the marginal distributions are specified as either Gaussian or t4 distributions. The parameters of the copulas are shown in Table \ref{tab:DGP_new}.

\begin{table}[htbp]
\centering
\caption{Summary of the four Data Generating Processes with different copula families.}
\label{tab:DGP_new}
\begin{tabular}{cccrcccccc}
\hline
\textbf{Tree} & \textbf{Edge} & \textbf{Conditioned} & \textbf{Conditioning}  & \textbf{Family}  & \textbf{Parameters} & \textbf{df} & \textbf{Tau} \\
\hline
1 & 1 & 3, 1 &          & gaussian & 0.34   & 1 & 0.218 \\
1 & 2 & 2, 1 &          & gaussian & 0.69   & 1 & 0.486 \\
2 & 1 & 4, 1 & 3        & gaussian & -0.046 & 1 & -0.029 \\
2 & 2 & 3, 2 & 1        & gaussian & 0.67   & 1 & 0.467 \\
3 & 1 & 4, 2 & 1, 3     & gaussian & -0.27  & 1 & -0.174 \\
\hline
\textbf{Tree} & \textbf{Edge} & \textbf{Conditioned} & \textbf{Conditioning} & \textbf{Family} & \textbf{Parameters} & \textbf{df} & \textbf{Tau} \\
\hline
1 & 1 & 3, 1 &          & clayton & 1.5  & 1 & 0.43 \\
1 & 2 & 2, 1 &          & clayton & 2.0  & 1 & 0.49 \\
2 & 1 & 4, 1 & 3        & clayton & 0.37 & 1 & 0.16 \\
2 & 2 & 3, 2 & 1        & clayton & 0.72 & 1 & 0.26 \\
3 & 1 & 4, 2 & 1, 3     & clayton & 0.24 & 1 & 0.11 \\
\hline
\textbf{Tree} & \textbf{Edge} & \textbf{Conditioned} & \textbf{Conditioning}  & \textbf{Family} &\textbf{Parameters} & \textbf{df} & \textbf{Tau} \\
\hline
1 & 1 & 3, 1 &          & frank & 2.0  & 1 & 0.214 \\
1 & 2 & 2, 1 &          & frank & 5.5  & 1 & 0.488 \\
2 & 1 & 4, 1 & 3        & frank & -0.57 & 1 & -0.063 \\
2 & 2 & 3, 2 & 1        & frank & 5.1  & 1 & 0.464 \\
3 & 1 & 4, 2 & 1, 3     & frank & -1.1 & 1 & -0.119 \\
\hline
\textbf{Tree} & \textbf{Edge} & \textbf{Conditioned} & \textbf{Conditioning} & \textbf{Family} & \textbf{Parameters} & \textbf{df} & \textbf{Tau} \\
\hline
1 & 1 & 3, 1 &          & joe  & 2.5 & 1 & 0.44 \\
1 & 2 & 2, 1 &          & joe  & 2.7 & 1 & 0.48 \\
2 & 1 & 4, 1 & 3        & joe  & 1.3 & 1 & 0.13 \\
2 & 2 & 3, 2 & 1        & joe  & 1.6 & 1 & 0.25 \\
3 & 1 & 4, 2 & 1, 3     & joe  & 1.2 & 1 & 0.08 \\
\hline
\end{tabular}
\end{table}

\begin{figure}
    \centering
    \begin{subfigure}{0.48\linewidth}
        \centering
        \includegraphics[width=\linewidth]{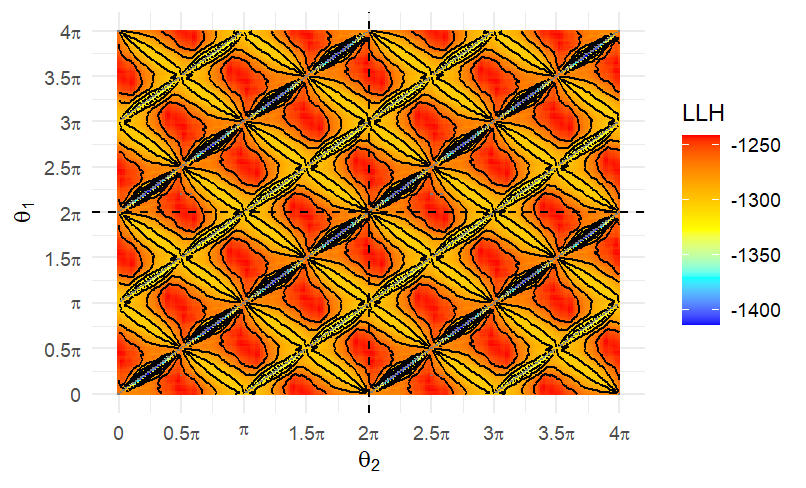}
        \caption{Gaussian copula}
        \label{fig:gaussian-t4}
    \end{subfigure}
    \hfill
    \begin{subfigure}{0.48\linewidth}
        \centering
        \includegraphics[width=\linewidth]{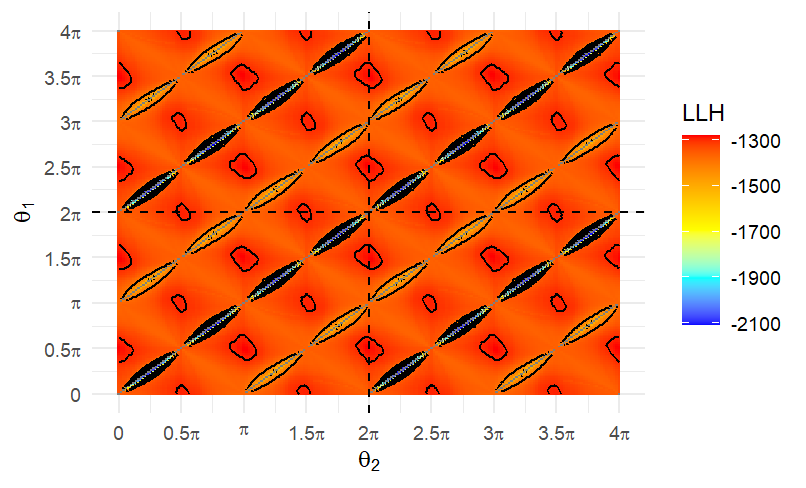}
        \caption{Frank copula}
        \label{fig:frank-t4}
    \end{subfigure}
    
    \begin{subfigure}{0.48\linewidth}
        \centering
        \includegraphics[width=\linewidth]{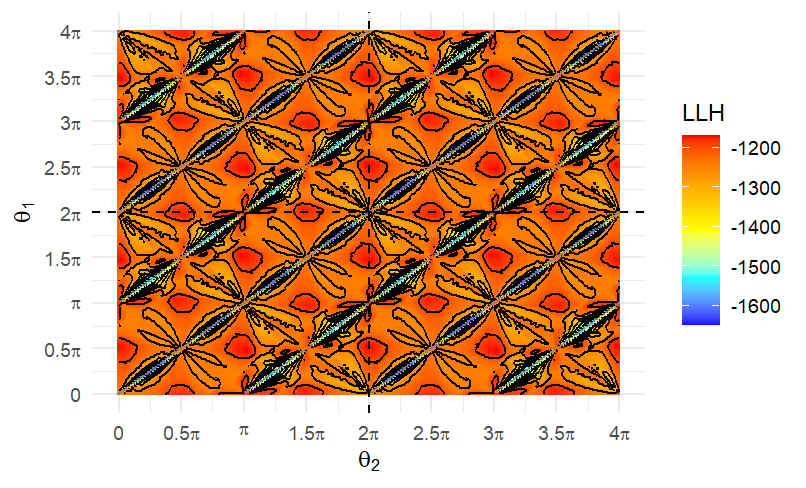}
        \caption{Clayton copula}
        \label{fig:clayton-t4}
    \end{subfigure}
    \hfill
    \begin{subfigure}{0.48\linewidth}
        \centering
        \includegraphics[width=\linewidth]{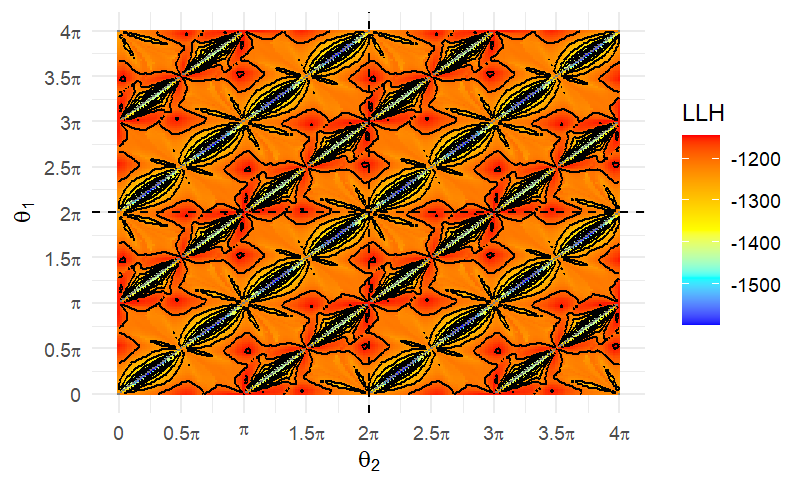}
        \caption{Joe copula}
        \label{fig:joe-t4}
    \end{subfigure}
    
    \caption{Log-likelihood contour plots after an affine transformation for different copula structures with \( t_4 \) marginals.}
    \label{fig3}
\end{figure}

\begin{figure}
    \centering
    \begin{subfigure}{0.48\linewidth}
        \centering
        \includegraphics[width=\linewidth]{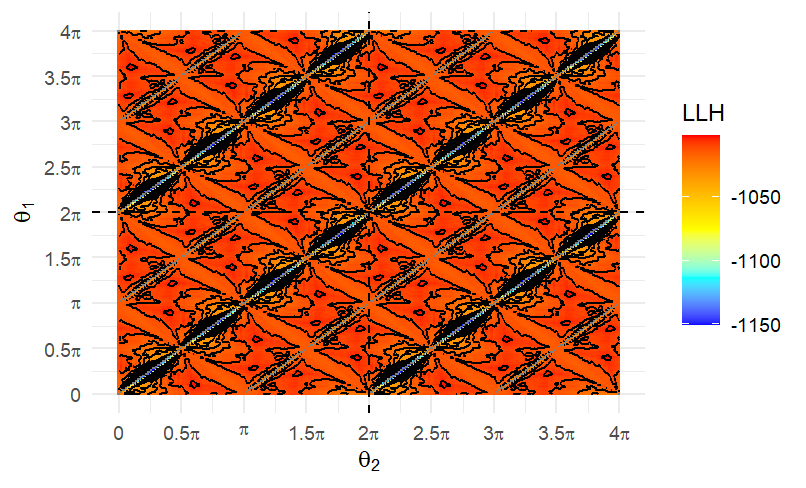}
        \caption{Gaussian copula}
        \label{fig:gaussian-gaussian}
    \end{subfigure}
    \hfill
    \begin{subfigure}{0.48\linewidth}
        \centering
        \includegraphics[width=\linewidth]{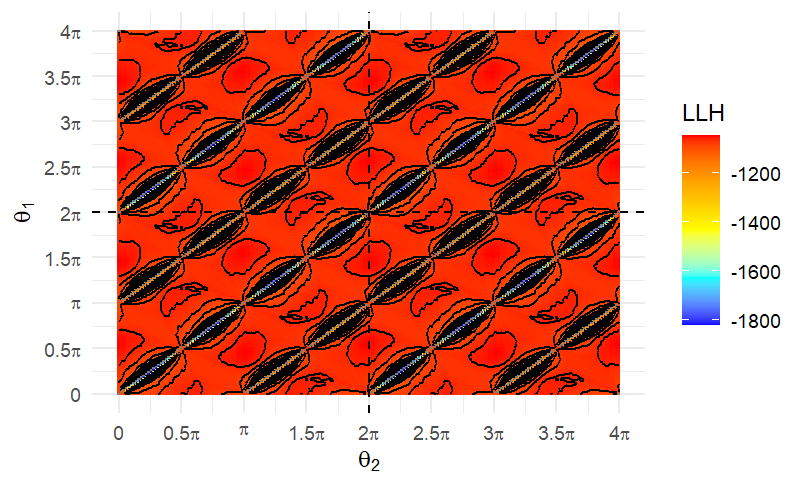}
        \caption{Frank copula}
        \label{fig:frank-gaussian}
    \end{subfigure}
    
    \begin{subfigure}{0.48\linewidth}
        \centering
        \includegraphics[width=\linewidth]{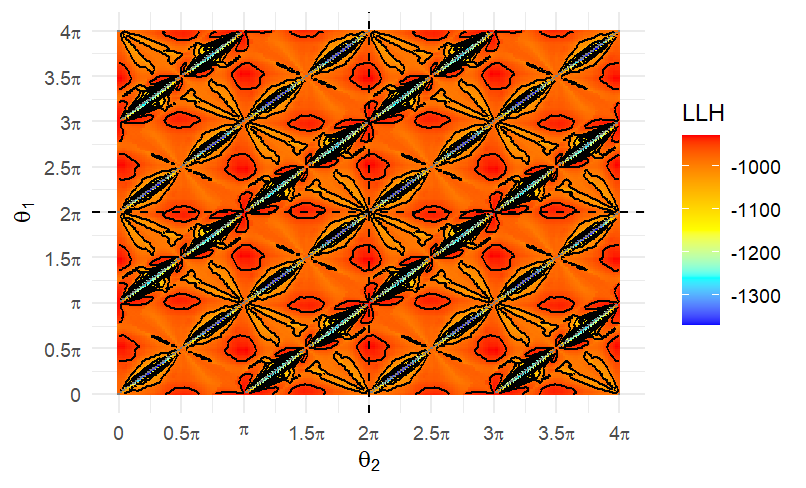}
        \caption{Clayton copula}
        \label{fig:clayton-gaussian}
    \end{subfigure}
    \hfill
    \begin{subfigure}{0.48\linewidth}
        \centering
        \includegraphics[width=\linewidth]{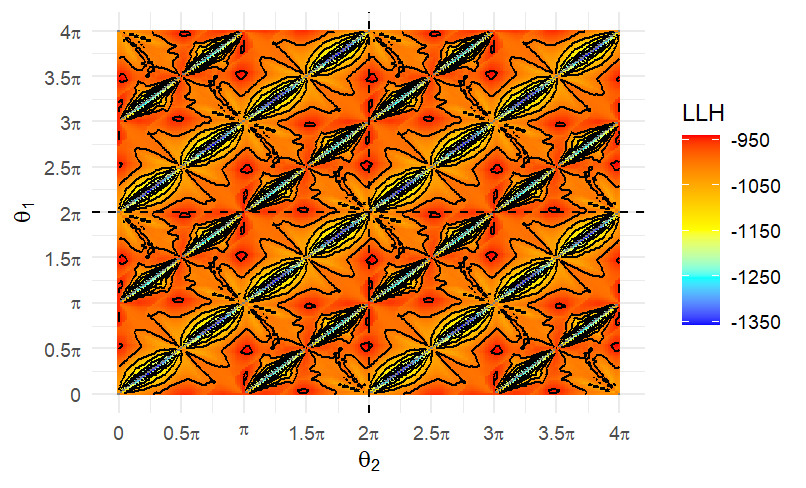}
        \caption{Joe copula}
        \label{fig:joe-gaussian}
    \end{subfigure}
    
    \caption{Log-likelihood contour plots after an affine transformation for different copula structures with standard normal marginals.}
    \label{fig4}
\end{figure}

The Figures \ref{fig3} and \ref{fig4} depict the log-likelihood (LLH) function contours after applying the affine transformation: \begin{equation*}
\begin{bmatrix}
\sin(\theta_{1}) & \sin(\theta_{2})\\
\cos(\theta_{1}) & \cos(\theta_{2})
\end{bmatrix},
\end{equation*}
for $\theta_1,\theta_2\in[0,4\pi]$, in order to reveal the periodic pattern.

Figure \ref{fig3} shows that when the marginal distributions are $t_4$, all the LLH contour plots achieve $(0.5\pi, 0)$, $(0, 0.5\pi)$ and have a period of $\pi$ in both directions.
This symmetry aligns with the estimation procedure’s allowance for copula rotations (e.g., 0$^{\circ}$, 90$^{\circ}$, 180$^{\circ}$, 270$^{\circ}$), as in \cite{NKM22}, which corresponds to sign-flipping transformations of factors.

Regarding the Gaussian marginal distributions, Clayton, Frank, and Joe copulas retain the LLH maxima at $(0.5\pi, 0)$ and $(0, 0.5\pi)$, but not for the Gaussian copula.  This discrepancy arises because rotating data in most case forces misspecification: the rotated joint distribution no longer resides within the original copula family, leading to inferior LLH values. The Gaussian copula with Gaussian margins, however, constitutes an exception. Affine transformations preserve the multivariate Gaussian structure, ensuring the copula family remains unchanged under rotation. 

Although for Gaussian data, we may not be able to identify the specific rotation due to the invariance properties of the multivariate Gaussian distribution, this underspecification does not imply misspecification. As a result, the model can still achieve strong forecasting performance despite the lack of identifiability. We demonstrate this with an simulated example in Section \ref{sec5.3}.

\subsection{Performance of estimators} \label{sec5.2}

In order to quantify the assessment of the estimations, we compute the Root Mean Square Error (RMSE) for the parameters, common factors, and factor loadings as follows

\begin{eqnarray}
&&\operatorname{RMSE}(\widehat{\boldsymbol \theta}_C)= \frac{1}{\sqrt{dim({\boldsymbol \theta}_C)}} \left\Vert \widehat{\boldsymbol \theta}_C-{\boldsymbol \theta}_C\right\Vert,\notag\\
&&\operatorname{RMSE}(\widehat{\boldsymbol{F}})=\left(\frac{1}{T}\sum_{t=1}^T\left\Vert {\bf H}(\widehat{\boldsymbol \theta}_H)\widehat{\boldsymbol{F}}_t-{\boldsymbol{F}}_t\right\Vert^2\right)^{1/2},\notag\\
&&\operatorname{RMSE}(\widehat\Lambda)=\left(\frac{1}{N}\sum_{i=1}^N\left\Vert {\bf H}(\widehat{\boldsymbol \theta}_H)^{-1}\widehat {\boldsymbol\lambda}_{i}- {\boldsymbol\lambda}_{i}\right\Vert^2\right)^{1/2},\notag
\end{eqnarray} 
where $ \dim({\boldsymbol \theta}_C) $ is the number of parameters in the vine copula. In practice, we sometimes obtain the flipping estimated factors, which has no influence on the same likelihood but leads to different estimates of the vine copulas and rotaion matrix parameters, ${\boldsymbol \theta}_C$ and ${\boldsymbol \theta}_H$. Therefore, we calculate the RMSEs based on the best sign-flipped estimates of the factors and the corresponding $\widehat{\boldsymbol \theta}_C$ and $\widehat{\boldsymbol \theta}_H$.

\bigskip

\begin{figure}[h!]
    \centering
    \includegraphics[width = 0.4\textwidth]{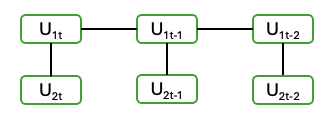}
    \caption{Vine structure of first tree of two-variable and second-order M-vine, where $U_{it}$ denotes the probability integral transform of $F_{it}$.}
    \label{fig:vs2}
\end{figure}

We now give a simulated example of an approximate factor model with two factors having a vine copula structure \citep[see][]{beare2015vine}. The vine structure is presented in Figure \ref{fig:vs2}, which is an M-vine of second-order. 
All pair copulas in the vine are Frank copulas and Table \ref{tab:DGP_summary} shows the specification of the M-vine copula model with the vine structure and associated parameters for each edge.

\begin{table}[htbp]
\centering
\caption{Summary of the Data Generating Process (DGP)}
\label{tab:DGP_summary}
\begin{tabular}{cccrcccccc}
\hline
\textbf{Tree} & \textbf{Edge} & \textbf{Conditioned} & \textbf{Conditioning}  & \textbf{Family}& \textbf{Parameters} & \textbf{df} & \textbf{Tau} \\
\hline
1 & 1 & 3, 1 &           & frank & 2 & 1 & 0.2168 \\
1 & 2 & 2, 1 &           & frank & 5.4 & 1 & 0.4781 \\
2 & 1 & 4, 1 & 3         & frank & -0.33 & 1 & -0.0361 \\
2 & 2 & 3, 2 & 1         & frank & 5 & 1 & 0.4540 \\
3 & 1 & 5, 1 & 4, 3      & frank & 0.16 & 1 & 0.0174 \\
3 & 2 & 4, 2 & 1, 3      & frank & -1.6 & 1 & -0.1772 \\
4 & 1 & 6, 1 & 4, 3, 5   & frank & -0.039 & 1 & -0.0043 \\
4 & 2 & 5, 2 & 1, 4, 3   & frank & 0.7 & 1 & 0.0776 \\
5 & 1 & 6, 2 & 1, 4, 3, 5& frank & 0.019 & 1 & 0.0021 \\
\hline
\end{tabular}
\end{table}

The common factors in the approximate factor model are generated by $\Phi^{-1}(U_{it})$, for $i=1,2$ and $t=1,\cdots,T$, where $\Phi(\cdot)$ is the c.d.f. of the standard normal distribution and $U_{it}$ are generated from the M-vine model with parameters specified in Table \ref{tab:DGP_summary}. Factor loadings are drawn independently from a normal distribution with mean and variance both equal to one. The idiosyncratic errors follow a zero-mean, unit-variance Gaussian AR(1) process with an autoregressive coefficient of 0.5. With the simulated factors, factor loadings, and idiosyncratic errors, we calculate the observations $X_{it}$ using the approximate factor model \eqref{eq3.1}, where the dimension $d$ is equal to 100, 200 or 500, and the number of observation $n$ is equal to 100, 250, 500, 750, 1000, or 2000. The simulations are repeated 200 times. The means of the RMSE for parameters, common factors, and factor loadings for the 200 simulations are presented in Table \ref{tab:RMSE of parameters}--\ref{tab:RMSE of Lambda}, respectively. 

\begin{table}[htbp] 
  \centering
  \caption{RMSE of Parameters.} 
  \begin{tabular}{*{1}{l}*{3}{r}}
    \toprule
    \( n \) \textbar\ d & \multicolumn{1}{c}{100} & \multicolumn{1}{c}{200} & \multicolumn{1}{c}{500} \\
    \midrule
     100 & 2.2413 & 2.2786 & 2.2471 \\
    250 & 1.8570 & 1.8471 & 1.8153 \\
    500 & 1.4506 & 1.4429 & 1.3591 \\
    750 & 1.2381 & 1.1697 & 1.1529 \\
    1000 & 1.0209 & 1.0308 & 0.8545 \\
    2000 & 0.6288 & 0.5736 & 0.6086 \\
    \bottomrule
  \end{tabular}
    \label{tab:RMSE of parameters}
\end{table}

\begin{table}[htbp]
  \centering
    \caption{RMSE of factors.} 
  \begin{tabular}{*{1}{l}*{7}{r}}
    \toprule
     &   \multicolumn{3}{c}{First Factor}  &   & \multicolumn{3}{c}{Second Factor} \\
    \hline
    \( n \) \textbar\ d & \multicolumn{1}{c}{100} & \multicolumn{1}{c}{200} & \multicolumn{1}{c}{500} &  & \multicolumn{1}{c}{100} & \multicolumn{1}{c}{200} & \multicolumn{1}{c}{500}\\
    \midrule
    100 & 0.4205 & 0.4217 & 0.4149 &  & 0.8435 & 0.8758 & 0.8320\\
    250 & 0.3460 & 0.3152 & 0.3412 &  & 0.7456 & 0.6995 & 0.7255\\
    500 & 0.2734 & 0.2602 & 0.2402 &  & 0.6317 & 0.6186 & 0.5704\\
    750 & 0.2501 & 0.2252 & 0.2119 &  & 0.5438 & 0.4978 & 0.4741\\
    1000 & 0.2100 & 0.1996 & 0.1636 &  & 0.4393 & 0.4381 & 0.3400\\
    2000 & 0.1588 & 0.1288 & 0.1216 &  & 0.2831 & 0.2368 & 0.2316\\
    \bottomrule
  \end{tabular}
  \label{tab:RMSE of factors}
\end{table}

\begin{table}[htbp]
  \centering
    \caption{RMSE of factor loadings. }
  \begin{tabular}{*{1}{l}*{7}{r}}
    \toprule
    &   \multicolumn{3}{c}{Lambda one}  &   & \multicolumn{3}{c}{Lambda two} \\
    \hline
    \( n \) \textbar\ d & \multicolumn{1}{c}{100} & \multicolumn{1}{c}{200} & \multicolumn{1}{c}{500} & & \multicolumn{1}{c}{100} & \multicolumn{1}{c}{200} & \multicolumn{1}{c}{500} \\
    \midrule
    100 & 1.0594 & 1.1097 & 1.0442 &  & 0.9976 & 1.0503 & 0.9663\\
    250 & 0.9195 & 0.8961 & 0.8978 &  & 0.8278 & 0.7884 & 0.8424\\
    500 & 0.7690 & 0.7663 & 0.7160 &  & 0.6727 & 0.6664 & 0.6249\\
    750 & 0.6651 & 0.6273 & 0.6091 &  & 0.5975 & 0.5557 & 0.5501\\
    1000 & 0.5270 & 0.5426 & 0.4430 &  & 0.4675 & 0.4840 & 0.3970\\
    2000 & 0.3182 & 0.2800 & 0.2959 &  & 0.2926 & 0.2539 & 0.2781\\
    \bottomrule
  \end{tabular}
  \label{tab:RMSE of Lambda}
\end{table}

The three tables show that the RMSEs of the estimated parameters, common factors, and factor loadings converge toward zero as both the dimension and the number of observations increase. Our method performs well even in high-dimensional settings where $n$ is smaller than $d$. On the one hand, the accuracy of factor estimation improves as the dimension increases, due to more information being available for identifying the common structure. On the other hand, a larger number of observations leads to more accurate parameter estimation in the second step. Overall, the number of observations  has a greater impact on estimation accuracy than the dimension.

\subsection{Forecasting performance} \label{sec5.3}

 In this section, we compare the one-step prediction performance of our S-vine factor models (SF) with Frank copula sequences to the SF model with mispecified pair copulas (Gaussian copulas), GARCH processes with skewed student t innovation distribution, and the dynamic factor model (DFM), where for forecasting using DFM see \cite{HT23}. Factors are generated from the two-variable second-order M-vine with pair copulas and parameters presented in Table \ref{tab:DGP_summary}. We forecast with the proposed SF model (\texttt{SF\_1}) and compare it with the SF model without factor rotation (\texttt{SF\_2}) and the SF model with a misspecified pair copula (\texttt{SF\_wrong}) uses Gaussian copulas as the selected copula family. With the kernel function, the rotation matrix, the simulated factor loadings, and the idiosyncratic component generated from a normal distribution, we simulate $X_{it}$. The dimension is set to 100 or 200 and the number of training data is 100, 250, 500 and 750, respectively. We set the number of test data to 200 for all these cases. 

In order to compare the accuracy of the distributional predictions, we calculate the quantile scores for each model. The score function proposed by \cite{gneiting2011comparing}, to assess the one-step prediction performance, which can be written as 
\begin{equation} \label{eq:score1}
    S_{\alpha}^Q(y,x) = |1_{\{y\leq x\}}-\alpha||y-x|,
\end{equation}
where $y$ is the forecast value at $\alpha$-quantile and $x$ is the realized value. Suppose that the one-step predictive distribution of $X_{it}$ is $\widehat{G}_{i,t+1}$ in S-vine factor models. We compute the one-step-ahead forecast of the Value-at-Risk $\widehat{VaR}_\alpha^t = \widehat{G}^{-1}_{i,t+1}(\alpha|\mathbf{X}_{it})$ at $\alpha$-quantile, for $t = 1,....,n$, and calculate the mean of quantile scores \citep[Chapter 9 in][]{mcneil2015quantitative}, 
\begin{equation}\label{eq:score2}
   \frac{1}{n}\sum_{t=1}^n S_{\alpha}^Q(\widehat{VaR}_\alpha^t, X_{it+1}).
\end{equation}
 In our simulation, we only present the results for $i=1$, as the results across dimensions are similar. 

The predictions are repeated 100 times. We calculate the average value of the mean of quantile scores for the 100 repetitions in each case and demonstrate it in Table \ref{tab:aqs}. According to Table \ref{tab:aqs}, the SF models have smaller average quantile scores than DFM and GARCH, which is reasonable, since the data are simulated from the SF models. \texttt{SF\_1} has smaller average quantile scores compares to SF\_2, which means the SF model with rotation matrix perform better in prediction. Therefore, the rotation is an effective step for SF models in the estimation and prediction process. 
The difference of the average quantile scores between the SF models with the "true" and "wrong" copulas is not pronounced, which implies that the type of copulas may not have significant influence on the SF models. 

\begin{table}[htbp]
  \centering
  \caption{Mean of quantile scores of 100 repetition of forecasting for two-variable and second-order factor M-vine with Frank copulas. The \texttt{SF\_1} model is the S-vine factor model we proposed.The \texttt{SF\_2} model is the S-vine factor model without rotation. The \texttt{SF\_wrong} is the the S-vine factor model with the same settings as the \texttt{SF\_1} model but with the Gaussian copulas as the pair copulas in M-vine. DFM represents the dynamic factor model.}
  \label{tab:aqs}
  \begin{tabular}{*{2}{l}*{5}{r}}
    \toprule
    \( n \) & \( d \) \textbar\ model & \multicolumn{1}{c}{SF\_1} & \multicolumn{1}{c}{SF\_2} & \multicolumn{1}{c}{SF\_wrong} & \multicolumn{1}{c}{DFM} & \multicolumn{1}{c}{GARCH} \\
    \midrule
    100 & 100 & 0.5370 & 0.5467 & 0.5525 & 0.7205 & 0.5871 \\
    & 200 & 0.5249 & 0.5356 & 0.5515 & 0.6956 & 0.5857 \\ \addlinespace[3pt]
    250 & 100 & 0.5320 & 0.5346 & 0.5359 & 0.7264 & 0.5899 \\
    & 200 & 0.5322 & 0.5354 & 0.5338 & 0.6858 & 0.5830 \\ \addlinespace[3pt]
    500 & 100 & 0.5175 & 0.5201 & 0.5302 & 0.7156 & 0.5778 \\
    & 200 & 0.5510 & 0.5525 & 0.5532 & 0.7278 & 0.6120 \\ \addlinespace[3pt]
    750 & 100 & 0.5268 & 0.5293 & 0.5363 & 0.7414 & 0.5907 \\
    & 200 & 0.5270 & 0.5298 & 0.5417 & 0.7050 & 0.5946 \\
    \bottomrule
  \end{tabular}
\end{table}

\begin{figure} 
    \centering
    \includegraphics[width=\linewidth]{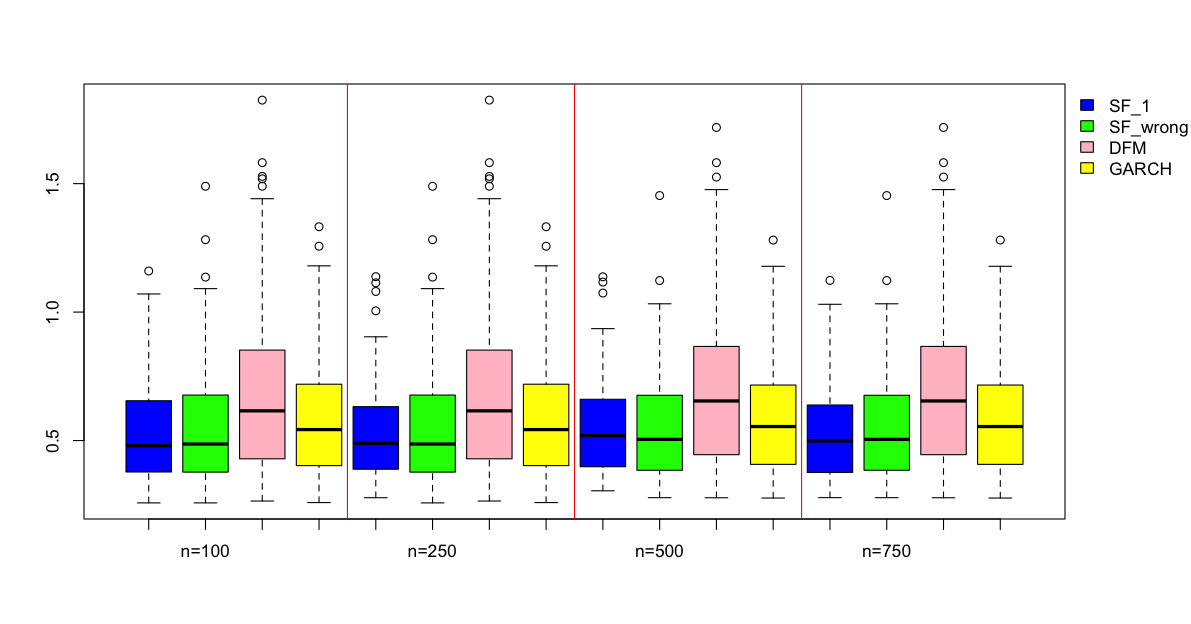}
    \caption{Boxplots of the 100 times repeated quantile scores of predictions from SF\_1, SF\_wrong, DFM and GARCH models, where the \texttt{SF\_1} model is the S-vine factor model we proposed. The \texttt{SF\_wrong} is the the S-vine factor model with the same settings as the \texttt{SF\_1} model but with the Gaussian copulas as the pair copulas in M-vine. DFM represents the dynamic factor model.}
    \label{fig:box}
\end{figure}

The box plots of the mean quantile scores across 100 simulations are shown in Figure \ref{fig:box}. We present results only for the case with \( d = 200 \). The \texttt{SF\_2} model is omitted from the plots due to its poor forecasting performance without rotation, as reported in Table \ref{tab:aqs}. The inadequacy of the incorrect SF model can be observed in the box plots, where the green box (representing \texttt{SF\_1}) is slightly higher than the blue box (representing the true model). Similarly, both the DFM and GARCH models exhibit higher quantile scores and greater variability compared to the SF models.

\section{Application}\label{sec6}

In this section, we apply the proposed method to estimate the dependence structure of the common factors driving the daily volatilities of the S\&P 500 Index constituents, and to predict the Value-at-Risk (VaR) of the daily return of the S\&P 500 Index. The return data are obtained from the London Stock Exchange Group database and cover the period from 1 January 2021 to 31 December 2024. We begin the sample in 2021 to avoid structural breaks in the S\&P 500 Index constituents, which occurred on 6 March 2020 and were detected by \cite{LLF23}. Missing values are filled by interpolation, and each stock’s returns are standardised to have zero mean and unit variance.
Following \cite{barigozzi2016generalized}, we adopt an approximate factor model for volatility, using the absolute values of returns as a proxy. To highlight the role of cross-sectional dependence and latent factors, we compare the predictive performance of our proposed factor copula model with that of a univariate benchmark. Specifically, we consider a GARCH(1,1) model with skewed Student-\emph{t} innovations \citep{bollerslev1986generalized}, which is a widely used and effective method for modelling financial return series.

The training set consists of data from the first three years (2021–2023), with the 2024 data reserved for one-step-ahead out-of-sample prediction. A total of 484 firms were listed on the S\&P 500 Index during the entire period. Additionally, we include the S\&P 500 Index itself as the 485th variable, with a particular focus on forecasting its Value-at-Risk (VaR). Specifically, we forecast the distribution of the absolute value of returns using the volatility factor model and assume equal probabilities for negative and positive returns to construct the VaR forecast.



\begin{figure}[h!]
    \centering
    \includegraphics[width = 0.7\textwidth]{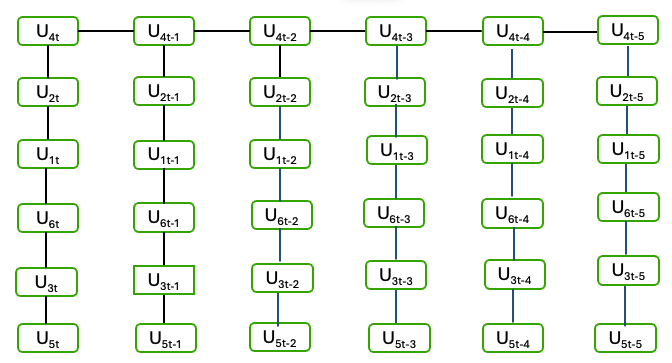}
    \caption{Vine structure of first tree of six-variable and fifth-order M-vine, where $U_{it}$ denotes the probability integral transform of $F_{it}$.}
    \label{fig:vs3}
\end{figure}

Let \( X_{it} \) denote the absolute stock return of asset \( i \) at time \( t \), where \( i = 1, \ldots, 485 \) and \( t = 1, \ldots, 753 \) corresponds to the training period. The remaining 252 observations are reserved for the test period. We employ a factor model with a fifth-order M-vine structure to fit the absolute returns, where the vine order is selected based on the autocorrelation function (ACF) plots of the estimated factors. The number of factors is determined using the information criterion proposed by \cite{BN02}, which identifies six latent factors.  The vine structure is shown in Figure \ref{fig:vs3}.

We fit the model using our proposed method under different pair copula families. The candidate pair copulas include the Gaussian, Frank, Clayton and "all" copulas, where the pair copulas in different edges are selected according to the AIC values. Table \ref{tab:LLK} shows the log-likelihood of the selected SF models. Using the estimated factor model with M-vine structure, we forecast the one-step-ahead distribution and calculate the Value-at-Risk (VaR) at 0.05, 0.10, 0.90, 0.95-quantile via an expanding window. The parameters and rotation matrix are fixed and without refit in the prediction process. 


\begin{table}

  \begin{minipage}[t]{0.4\textwidth}
    \centering
    \caption{Log-likelihood of SF models with Gaussian, Frank, Clayton and "all" copulas.}
  \label{tab:LLK}
  \begin{tabular}{*{1}{l}*{1}{r}}
    \toprule
    Family \textbar\ & Log-likelihood \\
    \midrule
    Gaussian & -5293.06 \\
    Frank & -5624.63 \\
    Clayton & -5301.46 \\
    all & -5011.98 \\
    \bottomrule
  \end{tabular}
    \end{minipage}
      \hfill
     \begin{minipage}[t]{0.52\textwidth}
    \centering
  \caption{Violations of the predicted VaR at 0.05, 0.01, 0.90, 0.95-quantile for SF models with Gaussian, Frank, Clayton and "all" copulas, and the GARCH(1,1) process with skewed t distribution. The "TRUE" represents the real violations. }
\label{tab:violation}
  \begin{tabular}{*{1}{l}*{4}{r}}
    \toprule
    Family \textbar\ Quantile & \multicolumn{1}{c}{0.05} & \multicolumn{1}{c}{0.10} & \multicolumn{1}{c}{0.90} & \multicolumn{1}{c}{0.95} \\
    \midrule
    Gaussian & 15 & 27 & 20 & 10 \\
    Frank & 13 & 27 & 21 & 9 \\
    Clayton & 12 & 25 & 19 & 11 \\
    all & 15 & 24 & 18 & 12 \\
    GARCH & 13 & 23 & 23 & 9 \\
    TRUE & 13 & 25 & 25 & 13 \\
    \bottomrule
  \end{tabular}
\end{minipage}
\end{table}

We calculate the violations of VaR of the one-step prediction in all these models and present it in Table \ref{tab:violation}.
The SF models with Frank and Clayton copulas have similar violations as GARCH process at 0.05 quantiles, while the SF with Gaussian and "all" copulas overestimate the violations. The GARCH process does not perform well at 0.95-quantile. On the contrary, the SF models with Clayton or "all" copulas show the advantages at 0.95-quantile. The violations at the lower and upper quantiles reveal the asymmetry of the left and right tail of the predictive distribution. For example, the SF models with Gaussian copulas overestimate the violations at lower quantile and underestimate the ones at upper quantile, which lead to understate of risk of loss and expectation of profit.The performance of SF models can be improved by selecting suitable combinations of pair copulas. The "all" copula model serves as a representative example, showing more precise violations across quantiles. 

\begin{figure}
    \centering
    \includegraphics[width=\linewidth]{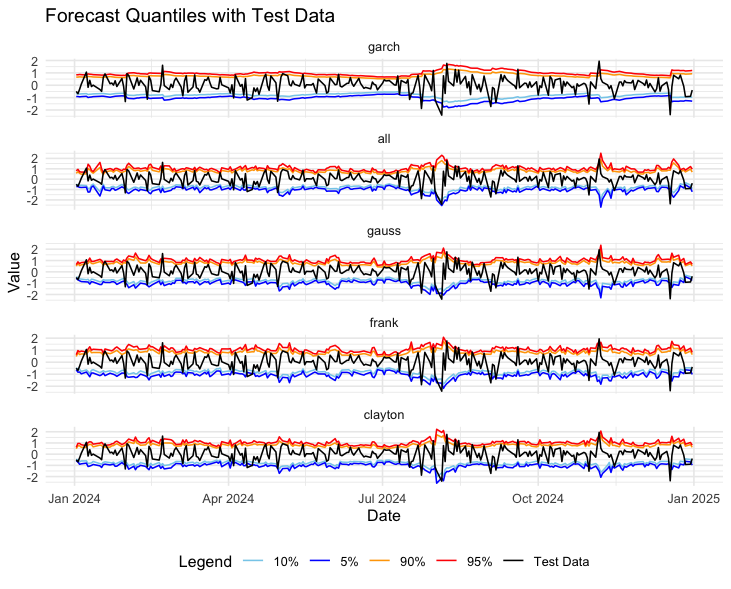}
    \caption{Forecast Quantiles at 0.05, 0.10, 0.90, 0.95 with S\&P500 index data. The black line represents the real returns in 2024. The light blue, blue, amber and red line represent forecast VaR at 0.01, 0.05, 0.90, 0.95-quantile, respectively.}
    \label{fig:FQ}
\end{figure}

 Figure \ref{fig:FQ} illustrates the forecasted quantiles at levels 0.05, 0.10, 0.90, and 0.95 for the GARCH model and SF models with the "all", Gaussian, Frank, and Clayton copulas, based on test data.
The VaR predicted by both the GARCH process and the SF models is influenced by extreme historical values. However, the GARCH model exhibits a smoother prediction, while the SF model displays a less persistent pattern and recovers to normal levels more quickly. Overall, the quantile plots on the test data show satisfactory performance.


\section{Conclusion}\label{sec7}
This paper introduced a novel framework by integrating an approximate factor model with an S-vine copula structure to provide extra flexibility and enhanced the precision of dependency modelling in high-dimensional data. We proposed a two-step estimation procedure and developed the one-step-ahead distributional prediction using the S-vine factor models. A factor rotation matrix was applied to the PCA factor estimates in order to accommodate their dependence and align them with a semiparametric S-vine copula model. The proposed rotation and the parameters of the S-vine copula are estimated via maximum likelihood.

We establish the consistency of rotation and parameter estimates in S-vine copulas. To achieve this, we develop asymptotic theory for the empirical process of latent factors under mixing conditions, thereby extending existing results from i.i.d. sequences to dependent settings. Furthermore, we derive uniform convergence results for kernel-based entropy estimators, explicitly accounting for the combined effects of projection, factor estimation error, and serial dependence. These theoretical justifications provide strong support for the validity and robustness of our proposed estimation framework.

The numerical experiments embodied the fact that the estimated parameters converge to the true parameters with the increase in dimensionality and sample size. We examined the forecasting performance using VaR estimation via Monte-Carlo methods in simulation studies and empirical applications. The practical identification issue is discussed and illustrated via simulation in Section \ref{sec4.1}, which sheds light on that the model can attain prominent and robust performance in one-step-ahead forecast process even when identification issues exists. We apply the daily returns of S\&P 500 Index constituents as an example to assess the forecasting performance of our model. The approximate factor model with S-vine structure demonstrated a pronounced advantage in VaR prediction, where it outperformed the GARCH process at various quantiles.

%
%
%
\bibliographystyle{elsarticle-harv} 
\bibliography{biblio}

\newpage

\appendix
\section{Proofs of the main results}\label{appendixA}
\renewcommand{\theequation}{A.\arabic{equation}}
\setcounter{equation}{0}
\renewcommand{\thelemma}{A.\arabic{lemma}}

\begin{lemma}\label{lemmaA.1}

{\em Suppose that Assumption \ref{as2}(i)--(vi) is satisfied. As $N$ and $T$ tend to infinity jointly,
\begin{equation}\label{eqA.2}
\left\{
\begin{array}{ll}
|\widetilde{v}_i-v_{i,0}|=O_P\left(T^{1/\delta_\varepsilon}N^{-1/2}+T^{-1/2}\right),&1\leq i\leq K,\\
\widetilde{v}_i=O_P\left(T^{1/\delta_\varepsilon}N^{-1/2}+T^{-1/2}\right),&K+1\leq i\leq N\wedge T,
\end{array}
\right.
\end{equation}
where $\widetilde{v}_i$ and $v_{i,0}$ denote the $i$-th largest eigenvalues of ${\bf X}{\bf X}^\top/NT$ and ${\boldsymbol\Sigma}_\Lambda^{1/2}{\boldsymbol\Sigma}_F{\boldsymbol\Sigma}_\Lambda^{1/2}$, respectively.}

\end{lemma}

\noindent{\bf Proof of Lemma \ref{lemmaA.1}}.\ 
Let $\boldsymbol{\varepsilon}_{\bullet i}=(\varepsilon_{1,i},\cdots,\varepsilon_{T,i})^\top$. By \eqref{eq3.2}, we readily have that
\begin{eqnarray}
{\boldsymbol\Omega}:=\frac{1}{N}{\bf X}{\bf X}^\top&=&{\bf F}\left[\frac{1}{N}\sum_{i=1}^N {\boldsymbol\lambda}_{i} {\boldsymbol\lambda}_{i}^\top\right]{\bf F}^\top+{\bf F}\left[\frac{1}{N}\sum_{i=1}^N{\boldsymbol\lambda}_{i}\boldsymbol{\varepsilon}_{\bullet i}^\top\right]+\left[\frac{1}{N}\sum_{i=1}^N \boldsymbol{\varepsilon}_{\bullet i}{\boldsymbol\lambda}_{i}^\top\right]{\bf F}^\top+\nonumber\\
&&\frac{1}{N}\sum_{i=1}^N (\boldsymbol{\varepsilon}_{\bullet i}\boldsymbol{\varepsilon}_{\bullet i}^\top-\operatorname{E}[\boldsymbol{\varepsilon}_{\bullet i}\boldsymbol{\varepsilon}_{\bullet i}^\top])+\frac{1}{N}\sum_{i=1}^N \operatorname{E}[\boldsymbol{\varepsilon}_{\bullet i}\boldsymbol{\varepsilon}_{\bullet i}^\top]\nonumber\\
&=:&{\boldsymbol\Omega}_1+{\boldsymbol\Omega}_2+{\boldsymbol\Omega}_3+{\boldsymbol\Omega}_4+{\boldsymbol\Omega}_5.\label{eqA.3}
\end{eqnarray}

We first prove that 
\begin{equation}\label{eqA.4}
\max_{1\leq i\leq q}\left\vert\widetilde{v}_i-v_i({\boldsymbol\Omega}_1/T)\right\vert=O_P\left(T^{1/\delta_{\varepsilon}}N^{-1/2}+T^{-1/2}\right).
\end{equation}
Following the argument in the proof of Lemma D.1 in \cite{LLF23}, we have
\begin{eqnarray}
\frac{1}{T}\Vert {\boldsymbol\Omega}_2\Vert=\frac{1}{T}\Vert {\boldsymbol\Omega}_3\Vert&=&\left\Vert{\bf F}\left[\frac{1}{NT}\sum_{i=1}^N {\boldsymbol\lambda}_{i}\boldsymbol{\varepsilon}_{\bullet i}^\top\right]\right\Vert\nonumber\\
&\leq&\left(\frac{1}{T}\sum_{t=1}^{T}\left\Vert {\boldsymbol F}_t\right\Vert^2\right)^{1/2}\left(\frac{1}{N^2T}\sum_{t=1}^T\left\Vert\sum_{i=1}^N{\boldsymbol\lambda}_{i}\varepsilon_{it}\right\Vert^2\right)^{1/2}\nonumber\\
&=&O_P\left(T^{1/\delta_\varepsilon}N^{-1/2}\right)=o_P(1),\label{eqA.5}
\end{eqnarray}
and 
\begin{eqnarray}
\frac{1}{T}\Vert {\boldsymbol\Omega}_4\Vert&=&\left\Vert\frac{1}{NT}\sum_{i=1}^N (\boldsymbol{\varepsilon}_{\bullet i}\boldsymbol{\varepsilon}_{\bullet i}^\top-\operatorname{E}[\boldsymbol{\varepsilon}_{\bullet i}\boldsymbol{\varepsilon}_{\bullet i}^\top])\right\Vert\nonumber\\
&\leq& \frac{1}{NT}\left(\max_{1\leq t \leq T}\sum_{s=1}^T\left(\sum_{i=1}^N \varepsilon_{t i}\varepsilon_{s i}-\operatorname{E}[ \varepsilon_{t i}\varepsilon_{s i}]\right)^2\right)^{1/2}\nonumber\\
&=&O_P\left(T^{2/\delta_\varepsilon-1/2}N^{-1/2}\right)=o_P(1).\label{eqA.6}
\end{eqnarray}
Note that a $\beta$-mixing process is also $\alpha$-mixing with $2\alpha_t\leq\beta_t$ \citep[e.g.,][pp.13--14]{LL97}. 
By a basic inequality on the covariance bound for the $\alpha$-mixing sequence \citep[e.g., Lemma 1.2.4 in][]{LL97}, we have
\begin{equation}\label{eqA.07}
\sum_{j=1}^{N}\left|\operatorname{E}\left[\varepsilon_{sj}\varepsilon_{tj}\right]\right|\leq 5\cdot\beta_{|s-t|}^{1-2/\delta_{\varepsilon}} \sum_{j=1}^{N} \left\{\mathsf{E}\left[|\varepsilon_{sj}|^{\delta_{\varepsilon}}\right]\right\}^{1/\delta_{\varepsilon}}\left\{\mathsf{E}\left[|\varepsilon_{tj}|^{\delta_{\varepsilon}}\right]\right\}^{1/\delta_{\varepsilon}}=O\left(N\cdot\beta_{|s-t|}^{1-2/\delta_{\varepsilon}}\right),
\end{equation}
indicating that
\begin{eqnarray}
\frac{1}{T}\Vert {\boldsymbol\Omega}_5\Vert&=&\left\Vert \frac{1}{NT}\sum_{i=1}^N \operatorname{E}[\boldsymbol{\varepsilon}_{\bullet i}\boldsymbol{\varepsilon}_{\bullet i}^\top]\right\Vert\nonumber\\
&\leq& \frac{1}{NT}\left(\max_{1\leq t \leq T}\sum_{s=1}^T\left(\sum_{i=1}^N \operatorname{E}[ \varepsilon_{t i}\varepsilon_{s i}]\right)^2\right)^{1/2}\nonumber\\
&=&O_P\left(T^{-1/2}\right)=o_P(1).\label{eqA.08}
\end{eqnarray}
By (\ref{eqA.3}), (\ref{eqA.5}), (\ref{eqA.6}), (\ref{eqA.08}) and Weyl's inequality, we complete the proof of (\ref{eqA.4}). As $\Vert\frac{1}{T}\sum_{t=1}^T {\boldsymbol F}_t{\boldsymbol F}_t^\top- {\boldsymbol \Sigma}_F\Vert=O_P(T^{-1/2})$ and $\Vert\frac{1}{N}\sum_{i=1}^N {\boldsymbol\lambda}_{i} {\boldsymbol\lambda}_{i}^\top-{\boldsymbol\Sigma}_\Lambda\Vert=O_P(N^{-1/2})$, we have
\begin{equation}\label{eqA.7}
\max_{1\leq i\leq q}\left\vert v_i({\boldsymbol\Omega}_1/T)-v_{i,0}\right\vert=O_P\left(N^{-1/2}+T^{-1/2}\right).
\end{equation}
Hence, by \eqref{eqA.4} and \eqref{eqA.7}, we prove $|\widetilde{v}_i-v_{i,0}|=O_P\left(T^{1/\delta_\varepsilon}N^{-1/2}+T^{-1/2}\right)$, for $1\leq i\leq K$. On the other hand, ${\boldsymbol\Omega}_1$ is a low-rank matrix with $\nu_i({\boldsymbol\Omega}_1)\equiv0$ when $i>K$. Then, with (\ref{eqA.5}) and (\ref{eqA.08}), we prove 
\[\widetilde{v}_i=O_P\left(T^{1/\delta_\varepsilon}N^{-1/2}+T^{-1/2}\right),\quad K+1\leq i\leq N\wedge T.
\]
The proof of Lemma \ref{lemmaA.1} is completed. \hfill$\Box$

\medskip

\noindent{\bf Proof of Proposition \ref{prop1}}.\
 By the definition of PCA estimation, we may show that
\begin{eqnarray}
{\boldsymbol V}_{NT}\left(\widehat{\boldsymbol{F}}_t-{\boldsymbol{R}_{NT}}{\boldsymbol{F}}_t\right)&=&\frac{1}{NT}\sum_{s=1}^T\sum_{j=1}^{N}\widehat{\boldsymbol{F}}_s{\boldsymbol{F}}_s^\top{\boldsymbol\lambda}_j \varepsilon_{tj}+\frac{1}{NT}\sum_{s=1}^T\sum_{j=1}^{N}\widehat{\boldsymbol{F}}_s{\boldsymbol{F}}_t^\top{\boldsymbol\lambda}_j\varepsilon_{sj}+\frac{1}{NT}\sum_{s=1}^T\sum_{j=1}^{N}\widehat{\boldsymbol{F}}_s\operatorname{E}\left[\varepsilon_{sj}\varepsilon_{tj}\right]\notag\\
&&+\frac{1}{NT}\sum_{s=1}^T\sum_{j=1}^{N}\widehat{\boldsymbol{F}}_s\left\{\varepsilon_{sj}\varepsilon_{tj}-\operatorname{E}\left[\varepsilon_{sj}\varepsilon_{tj}\right]\right\}\notag\\
&=:&{\boldsymbol\Pi}_{F1,t}+{\boldsymbol\Pi}_{F2,t}+{\boldsymbol\Pi}_{F3,t}+{\boldsymbol\Pi}_{F4,t}
\label{eqD.3}
\end{eqnarray}
for any $1\leq t\leq T$. \cite{LLF23} proves that $\max_{1\leq t\leq T}\Vert{\boldsymbol \Pi}_{F3,t}\Vert=O_P(T^{-1/2})$ and that $\max_{1\leq t\leq T}\Vert{\boldsymbol \Pi}_{Fi,t}\Vert=O_P(T^{2/\delta_\varepsilon}N^{-1/2})$, for $i=1,2,4$. By Lemma \ref{lemmaA.1}, ${\boldsymbol V}_{NT}$ converge to a matrix with bounded eigenvalues, and thus we have
\begin{equation}\label{eqA.010}
\max_{1\leq t\leq T}\left\Vert\widehat{\bf{F}}_t-{\bf R}_{NT}{\bf{F}}_t\right\Vert=O_P\left(\frac{1}{T^{1/2}}+\frac{T^{2/\delta_{\varepsilon}}}{N^{1/2}}\right)=o_p(1).
\end{equation}
We can further improve the uniform convergence rate by improving the bound for $\max_{1\leq t\leq T} \Vert{\boldsymbol\Pi}_{F3,t}\Vert_2$.
By the sub-Gaussian moment condition on ${\mathbf F}_t$ in Assumption \ref{as1}(iii), the Bonferroni and Markov inequalities, we have $\max_{1\leq t\leq T}\Vert{\mathbf F}_t \Vert=O_P\left((\log T)^{1/2}\right)$. By the H\"older inequality, \eqref{eqA.07} and \eqref{eqA.010}, we have 
\begin{eqnarray}
\max_{1\leq t\leq T} \Vert{\boldsymbol\Pi}_{F3,t}\Vert_2&=&\frac{1}{NT}\cdot\max_{1\leq t\leq T}\left\Vert\sum_{s=1}^T\sum_{j=1}^{N}\widehat{\boldsymbol{F}}_s\mathsf{E}\left[\varepsilon_{sj}\varepsilon_{tj}\right]\right\Vert_2\notag\\
&\leq&\frac{1}{NT}\cdot\left(\sum_{j=1}^K\max_{1\leq s\leq T}\widehat{F}_{sj}^2\right)^{1/2}\cdot\max_{1\leq t\leq T}\sum_{s=1}^T\left|\sum_{j=1}^{N}\mathsf{E}\left[\varepsilon_{sj}\varepsilon_{tj}\right]\right|\notag\\
&=&\frac{1}{NT}\cdot O_P\left((\log T)^{1/2}\right)\cdot O\left(N\cdot\sum_{k=1}^T\beta_k^{1-2/\delta_\varepsilon}\right)\notag\\
&=&O_P\left((\log T)^{1/2}T^{-1}\right),\label{eqD.8}
\end{eqnarray}
as $\sum_{k=1}^T\beta_k^{1-2/\delta_\varepsilon}<\infty$ when $\beta_k$ decays to zero at a geometric rate. Therefore, 
\begin{equation}
\max_{1\leq t\leq T}\left\Vert\widehat{\bf{F}}_t-{\bf R}_{NT}{\bf{F}}_t\right\Vert=O_P\left(\frac{(\log T)^{1/2}}{T}+\frac{T^{2/\delta_{\varepsilon}}}{N^{1/2}}\right).\label{eqA.013}
\end{equation}

Next, we prove \eqref{eq4.3}. Let
\[
{\boldsymbol\Sigma}_{\Lambda,N}=\frac{1}{N}\sum_{i=1}^N{\boldsymbol\lambda}_{i}{\boldsymbol\lambda}_{i}^\top,\ \ {\boldsymbol\Sigma}_{F,T}=\frac{1}{T} {\boldsymbol F}^\top{\boldsymbol F},\ \ \widetilde{\boldsymbol\Sigma}_{F,T}=\frac{1}{T} {\boldsymbol F}^\top\widetilde{\boldsymbol F},
\]
and 
\[\widetilde{\bf W}_{NT}={\bf W}_*{\boldsymbol D}_W^{-1},\ \ {\bf W}_*= {\boldsymbol\Sigma}_{\Lambda,N}^{1/2}\widetilde{\boldsymbol\Sigma}_{G,T},\ \ \ {\boldsymbol D}_{W_*}=\left(\operatorname{diag}\left\{{\bf W}_*^\top{\bf W}_*\right\}\right)^{1/2},\]
where $\operatorname{diag}\{\cdot\}$ denotes the diagonalisation of a square matrix. Write
\[
{\boldsymbol\Delta}_{NT}= {\boldsymbol\Sigma}_{\Lambda,N}^{1/2}{\boldsymbol\Sigma}_{G,T}{\boldsymbol\Sigma}_{\Lambda,N}^{1/2},\ \ \ 
{\boldsymbol\Delta}_{\ast}={\boldsymbol\Sigma}_{\Lambda,N}^{1/2}{\bf F}^\top\left(\frac{1}{T}{\boldsymbol\Omega}-\frac{1}{T}{\boldsymbol\Omega}_1\right)\widetilde 
{\bf F},
\]
where ${\boldsymbol\Omega}$ and ${\boldsymbol\Omega}_1$ are defined in (\ref{eqA.3}). 
It follows from the definition of the PCA estimation that
\begin{equation}\label{eqA.8}
\left({\boldsymbol\Delta}_{NT}+{\boldsymbol\Delta}_{\ast} {\bf W}_*^{-1}\right)\widetilde{\bf W}_{NT}=\widetilde{\bf W}_{NT}{\boldsymbol V}_{NT}.
\end{equation}
Hence $\widetilde{\bf W}_{NT}$ consists of the eigenvectors of ${\boldsymbol\Delta}_{NT}+{\boldsymbol\Delta}_{\ast} {\bf W}_*^{-1}$. Write 
\[{\boldsymbol R}_{NT}={\boldsymbol V}_{NT}^{-1}\left({\boldsymbol D}_W\widetilde{\bf W}_{NT}^\top\right){\boldsymbol\Sigma}_{\Lambda,N}^{1/2}.\]
With the triangle inequality, 
\begin{eqnarray}
\left\Vert{\boldsymbol R}_{NT}-{\boldsymbol R}_0 \right\Vert
&\leq&\left\Vert\left({\boldsymbol V}_{NT}^{-1}-{\boldsymbol V}_0^{-1}\right){\boldsymbol D}_{W_*}\widetilde{\boldsymbol W}_{NT}^{^\intercal}{\boldsymbol\Sigma}_{\Lambda,N}^{1/2}\right\Vert+\left\Vert{\boldsymbol V}_0^{-1}\left({\boldsymbol D}_{W_*}-{\boldsymbol V}_0^{1/2}\right)\widetilde{\boldsymbol W}_{NT}^{^\intercal}{\boldsymbol\Sigma}_{\Lambda,N}^{1/2}\right\Vert+\nonumber\\
&&\left\Vert{\boldsymbol V}_0^{-1/2}\left(\widetilde{\boldsymbol W}_{NT}-{\boldsymbol W}_0\right)^{^\intercal}{\boldsymbol\Sigma}_{\Lambda,N}^{1/2}\right\Vert+\left\Vert{\boldsymbol V}_0^{-1/2}{\boldsymbol W}_0^{^\intercal}\left({\boldsymbol\Sigma}_{\Lambda,N}^{1/2}-{\boldsymbol \Sigma}_\Lambda^{1/2}\right)\right\Vert,\label{eqA.9}
\end{eqnarray}
to complete the proof of (\ref{eq4.3}), we only need to show 
\begin{eqnarray}
\left\Vert {\boldsymbol V}_{NT}-{\boldsymbol V}_0\right\Vert&=&O_P\left(T^{1/\delta_{\varepsilon}}N^{-1/2}+T^{-1/2}\right),\label{eqA.10}\\
\left\Vert {\boldsymbol D}_W^2-{\boldsymbol V}_0\right\Vert&=&O_P\left(T^{1/\delta_{\varepsilon}}N^{-1/2}+T^{-1/2}\right),\label{eqA.11}\\
\left\Vert \widetilde{\bf W}_{NT}-{\bf W}_0\right\Vert&=&O_P\left(T^{1/\delta_{\varepsilon}}N^{-1/2}+T^{-1/2}\right),\label{eqA.12}\\
\left\Vert{\boldsymbol\Sigma}_{\Lambda,N}-{\boldsymbol \Sigma}_\Lambda\right\Vert&=&O(N^{-1/2}).\label{eqA.13}
\end{eqnarray}
By Lemma \ref{lemmaA.1}, we have  (\ref{eqA.10}). With the triangle inequality, 
\[
\left\Vert {\boldsymbol D}_W^2-{\boldsymbol V}_0\right\Vert\leq \left\Vert {\boldsymbol D}_W^2-{\boldsymbol V}_{NT}\right\Vert+\left\Vert {\boldsymbol V}_{NT}-{\boldsymbol V}_0\right\Vert
\]
and noting that
\begin{eqnarray}
\left\Vert {\boldsymbol D}_W^2-{\boldsymbol V}_{NT}\right\Vert&\leq&\left\Vert \frac{1}{T}\widetilde{\bf F}^\top\left(\frac{1}{T}{\boldsymbol\Omega}-\frac{1}{T}{\bf F} {\boldsymbol\Sigma}_{\Lambda,N}{\bf F}^\top\right)\widetilde{\bf F}\right\Vert\notag\\
&=&\left\Vert\frac{1}{T}{\boldsymbol\Omega}-\frac{1}{T}{\bf F} {\boldsymbol\Sigma}_{\Lambda,N}{\bf F}^\top\right\Vert\notag\\
&=&\frac{1}{T}\left\Vert {\boldsymbol\Omega}-{\boldsymbol\Omega}_1\right\Vert=O_P\left(T^{1/\delta_{\varepsilon}}N^{-1/2}+T^{-1/2}\right),\notag
\end{eqnarray}
using (\ref{eqA.4}) in the proof of Lemma \ref{lemmaA.1}, we can prove \eqref{eqA.11}. We next turn to the proof of \eqref{eqA.12}. 
By (\ref{eqA.8}), Lemma \ref{lemmaA.1} and the $\sin \theta$ theorem in \cite{DK70}, we have
\[
\left\Vert \widetilde{\bf W}_{NT}-{\bf W}_{NT}\right\Vert\leq C\cdot \Vert {\boldsymbol\Delta}_{NT}+{\boldsymbol\Delta}_{\ast}{\bf W}_{\ast}^{-1}-{\boldsymbol\Delta}_{0}\Vert.
\]
By Assumption \ref{as2}(ii)(iii), we have $\Vert {\boldsymbol\Delta}_{NT}-{\boldsymbol\Delta}_{0}\Vert=O_P\left(N^{-1/2}+T^{-1/2}\right)$.
Using Assumption \ref{as2}(ii)(iii) again and noting that the rotation matrix ${\boldsymbol R}_{NT}$ is asymptotically non-singular, we may show that $\Vert {\bf W}_{\ast}^{-1}\Vert=O_P(1)$. Following the proof of Lemma \ref{lemmaA.1}, we readily have $\Vert {\boldsymbol\Delta}_{\ast}\Vert=O_P\left(T^{1/\delta_{\varepsilon}}N^{-1/2}+T^{-1/2}\right)$, which leads to (\ref{eqA.13}). Combining (\ref{eqA.9})--(\ref{eqA.13}), the proof of \eqref{eq4.3} is completed.

To prove \eqref{eq4.4}
By the sub-Gaussian moment condition on ${\mathbf F}_t$ in Assumption \ref{as1}(iii), the Bonferroni and Markov inequalities, we have $\max_{1\leq t\leq T}\Vert{\mathbf F}_t \Vert=O_P\left((\log T)^{1/2}\right)$, which together with (\ref{eq4.3}), leads to 
\begin{eqnarray}
\max_{1\leq t\leq T}\left\Vert\widehat{\bf{F}}_t-{\bf R}_0{\bf{F}}_t\right\Vert&\leq&\max_{1\leq t\leq T}\left\Vert\widehat{\bf{F}}_t-{\bf R}_{NT}{\bf{F}}_t\right\Vert+\Vert{\bf R}_{NT}-{\bf R}_0\Vert\cdot\max_{1\leq t\leq T}\left\Vert{\bf{F}}_t\right\Vert\nonumber\\
&=&O_P\left((\log T)^{1/2}\left[\frac{1}{T^{1/2}}+\frac{T^{2/\delta_{\varepsilon}}}{N^{1/2}}\right]\right).\nonumber
\end{eqnarray}
Thus, we complete the proof of Proposition \ref{prop1}.
\hfill$\Box$

\medskip
\noindent{\bf Proof of Theorem \ref{thm4.1}}.\ Define $z_{\boldsymbol{\theta},t}={\boldsymbol H}_j^\top(
\widehat{\boldsymbol F}_t-{\bf R}_0{\boldsymbol F}_t)$. By the triangle inequality, We have 
\begin{eqnarray}
&&\left|\widehat G_{j}(x;{\boldsymbol \theta}_H)-\widetilde G_j(x;{\boldsymbol \theta}_H)+\frac{1}{T}\sum_{t=1}^Tg_{jt}(x;{\boldsymbol \theta}_H)z_{{\boldsymbol \theta}_H t}\right|\nonumber\\
&\leq&\left|\frac{1}{T}\sum_{t=1}^T\left({\boldsymbol 1}({\boldsymbol H}_j^\top \widehat{\boldsymbol F}_t\leq x)-G_{j}(\widehat{\boldsymbol F}_t;{\boldsymbol \theta}_H)-{\boldsymbol 1}({\boldsymbol H}_j^\top {\bf R}_0{\boldsymbol F}_t\leq x)+ G_{j}({\bf R}_0{\boldsymbol F}_t;{\boldsymbol \theta}_H)\right)\right|
+\nonumber\\
&&\left|\frac{1}{T}\sum_{t=1}^T\left(G_{j}(\widehat{\boldsymbol F}_t;{\boldsymbol \theta}_H)-G_{j}({\bf R}_0{\boldsymbol F}_t;{\boldsymbol \theta}_H)+g_{jt}(x;{\boldsymbol \theta}_H)z_{{\boldsymbol \theta}_H t}\right)\right|\nonumber\\
&:=&{\boldsymbol\Pi}_{G,1}+{\boldsymbol\Pi}_{G,2}.\label{eqA.20}
\end{eqnarray}
By Lemma \ref{lemmaB.4}, we have ${\boldsymbol\Pi}_{G,1}=o_P(T^{-1/2})$. By Assumption \ref{as2}(vii), Proposition \ref{prop1}, and the Taylor expansion, we can prove
\begin{eqnarray*}
{\boldsymbol\Pi}_{G,2}
&\leq&\sup_{{\boldsymbol \theta}_H \in {\boldsymbol \Theta}_H}\sup _{x \in \mathbb{R}}g_j'(x;{\boldsymbol \theta}_H)\cdot \frac{1}{T}\sum_{t=1}^Tz_{{\boldsymbol \theta}_H t}^2=o_P(T^{-1/2}),
\end{eqnarray*}
which, together with \eqref{eqA.20}, completes the proof of \eqref{eq4.5}.

Then by Lemma \ref{lemmaB.1}, we can prove \eqref{eq4.6}, Thus we complete the proof of Theorem \ref{thm4.1}.
\hfill$\Box$

\medskip
Before we prove Theorem \ref{thm4.2}, we introduce some notations.
 Assume the absolute regular process $\{\boldsymbol{F}_t\}$ 
 and the associated short notation 
 $$\mathbb{P} m=\operatorname{E}\left[m\left({\bf R}_0\boldsymbol{F}_t\right)\right], \mathbb{P}_T m=\frac{1}{T} \sum_{t=1}^T m\left({\bf R}_0\boldsymbol{F}_t\right),\ \ \textit{ and }\ \ \widehat{\mathbb{P}}_T m=\frac{1}{T} \sum_{t=1}^T m\left(\widehat{\boldsymbol{F}}_t\right),$$
 for expectation and the empirical measures over an arbitrary measurable function $m$.

Suppose that we want to estimate a parameter $\boldsymbol{\theta}^*=((\boldsymbol{\theta}_1^*)^\top,(\boldsymbol{\theta}_2^*)^\top)^\top\in \boldsymbol{\Theta}_1\times \boldsymbol{\Theta}_2=\boldsymbol{\Theta}$, where $\boldsymbol{\Theta}$ is a compact set in a finite-dimensional vector space equipped with a Euclidean norm in the presence of a nuisance parameter $\boldsymbol{\nu}$ in a function space equipped with a uniform norm. 
Specifically, we assume that, for any ${\boldsymbol{\theta}_1}$,  $\boldsymbol{\nu}({\boldsymbol{\theta}_1}):=\boldsymbol{\nu}({\boldsymbol{\theta}_1},\cdot)$ is a vector of functions on $\mathbb{R}$. 
Let $\widehat{\boldsymbol{\nu}}=\widehat {\boldsymbol{G}} - \boldsymbol{G}$ and $\boldsymbol{\nu}^*=\boldsymbol{0}$ (a vector of zero functions).
Denote $\left\|\boldsymbol{\nu}-\boldsymbol{\nu}_2\right\|$ as $\sup_{\boldsymbol{\theta}_1\in\boldsymbol{\Theta}_1}\sup_{x \in \mathbb{R}}\left\|\boldsymbol{\nu}(\boldsymbol{\theta}_1,x)-\boldsymbol{\nu}_2(\boldsymbol{\theta}_1,x)\right\|$. 
Let $\boldsymbol{\phi}_{\boldsymbol{\theta},\boldsymbol{\nu}(\boldsymbol{\theta}_1)}$ be a vector of function such that $\boldsymbol{\theta}^*$ solves $\mathbb{P}\boldsymbol{\phi}_{\boldsymbol{\theta},\boldsymbol{\nu}^*(\boldsymbol{\theta}_1)}=\boldsymbol 0$ and $\widehat{\boldsymbol{\theta}}$ solves $\mathbb{P}_T\boldsymbol{\phi}_{{\boldsymbol{\theta}},\widehat{\boldsymbol{\nu}}({\boldsymbol{\theta}_1})}=\boldsymbol 0$, where $\widehat{\boldsymbol{\nu}}(\boldsymbol{\theta}_1)$ is an estimator of ${\boldsymbol{\nu}}^*(\boldsymbol{\theta}_1)$.  We shall assume that $\cal{R}$ is a subset of a Banach space and define
$${\cal A}(\delta)= \{\boldsymbol{\theta}:\Vert\boldsymbol{\theta}-\boldsymbol{\theta}^*\Vert\leq\delta \}\ \ \textit{ and }\ \ {\cal R}(\delta)=\{\boldsymbol{\nu}:\Vert\boldsymbol{\nu}-\boldsymbol{\nu}^*(\boldsymbol{\theta}_1^*)\Vert\leq\delta \}.$$

\medskip
\noindent{\bf Proof of Theorem \ref{thm4.2}}.\ \
With the similar notation as in \eqref{eq4.4} and \eqref{eq4.7}, we can write $\widehat{\boldsymbol \theta}$ as the solution to 
\begin{equation}
\frac{1}{T}\sum_{t=1}^T{\boldsymbol{\phi}}_{\boldsymbol \theta,\widehat{\boldsymbol{G}}}(\widehat{\boldsymbol F}_t,\cdots,\widehat{\boldsymbol  F}_{t+p})+\widehat{\boldsymbol \psi}_{{\boldsymbol \theta}_H}=\boldsymbol{0},
\end{equation}
with convention that ${\boldsymbol \phi}_{{\boldsymbol \theta},\boldsymbol {\nu}}({\boldsymbol x}_t,\cdots,{\boldsymbol x}_{t+p})=\boldsymbol{0}$ when $t>T-p$,
where
\begin{equation}
\widehat{\boldsymbol \psi}_{{\boldsymbol \theta}_H}=\left(
\begin{aligned}
&\boldsymbol 0 \\
&\boldsymbol{s}_{1,{\boldsymbol \theta}_H}+ \widehat{\boldsymbol{s}}_{2,{\boldsymbol \theta}_H}
\end{aligned}\right).
\end{equation}
Because $\widehat{\boldsymbol \theta}=\argmin_{{\boldsymbol \theta}\in{\boldsymbol \Theta}} \left\Vert\widehat{\mathbb{P}}_T{\boldsymbol{\phi}}_{\boldsymbol{\theta},\widehat{\boldsymbol{G}}}+\widehat{\boldsymbol{\psi}}_{{\boldsymbol \theta}_H}\right\Vert$ and  ${\boldsymbol \theta}^*=\argmin_{{\boldsymbol \theta}\in{\boldsymbol \Theta}} \left\Vert\mathbb{P}{\boldsymbol{\phi}}_{\boldsymbol{\theta},{\boldsymbol{G}}}+{\boldsymbol{\psi}}_{{\boldsymbol \theta}_H}\right\Vert$, 
following from the Argmax theorem \citep[][Theorem 3.2.2]{VW96}, we only need to prove that
\begin{equation}\label{eqA.18}
\sup_{{\boldsymbol \theta}\in {\boldsymbol \Theta}} \left\Vert{\mathbb{P}}_T\boldsymbol{\phi}_{\boldsymbol{\theta},\widehat{\boldsymbol{G}}}-\mathbb{P}\boldsymbol{\phi}_{\boldsymbol{\theta},\boldsymbol{G}}\right\Vert=o_P(1),
\end{equation}
\begin{equation}\label{eqA.19}
\sup_{{\boldsymbol \theta}\in {\boldsymbol \Theta}} \left\Vert\widehat{\mathbb{P}}_T{\boldsymbol{\phi}}_{\boldsymbol{\theta},\widehat{\boldsymbol{G}}}-\mathbb{P}_T\boldsymbol{\phi}_{\boldsymbol{\theta},\widehat{\boldsymbol{G}}}\right\Vert=O_P\left(\frac{1}{T^{1/2}}+\frac{T^{2/\delta_{\varepsilon}}}{N^{1/2}}\right),
\end{equation}
and 
\begin{equation}\label{eqA.26}
\sup_{{\boldsymbol \theta}_H\in {\boldsymbol \Theta}_H}\left\Vert\widehat{\boldsymbol \psi}_{{\boldsymbol \theta}_H}-{\boldsymbol \psi}_{{\boldsymbol \theta}_H}\right\Vert=o_P(1),
\end{equation}

To prove \eqref{eqA.18}, by Theorem A.1 of \cite{NKM22}, we only need to check the conditions as follows,
\begin{itemize}
    \item[(C1)] The series $\left(\boldsymbol{F}_t\right)_{\mathrm{t} \in \mathrm{Z}}$ is strictly stationary and absolutely regular.
    \item [(C2)] For any $\delta>0$,
    \begin{equation}\label{eqA.21}
    \mathrm{P}\left(\left\Vert \widehat{\boldsymbol{G}} - \boldsymbol{G}\right\Vert  \leq \delta\right) \rightarrow 1
    \end{equation}
            as $N,T \rightarrow \infty$ and $b\rightarrow 0$.

    \item [(C3)] For every $\epsilon>0$, it holds $\inf_{\left\Vert {\boldsymbol{\theta}}-{\boldsymbol{\theta}}^*\right\Vert >\epsilon}\left\Vert \mathbb{P} {\boldsymbol \phi}_{{\boldsymbol{\theta}},\boldsymbol{G}} \right\Vert>0$.
    \item [(C4)] It holds $\mathbb{P}\sup _{{\boldsymbol{\theta}} \in {\boldsymbol{\Theta}}}\left\Vert {\boldsymbol \phi}_{{\boldsymbol{\theta}}, \boldsymbol{G}}\right\Vert <\infty$ and there is $\delta>0$ such that

$$P\left\{\sup _{{\boldsymbol{\theta},{\boldsymbol{\theta}'}\in{\boldsymbol{\Theta}}}} \sup _{\boldsymbol{\nu} , \boldsymbol{\nu}' \in \mathcal{R}(\delta)} \frac{\left\Vert {\boldsymbol \phi}_{{\boldsymbol{\theta}}_1, \boldsymbol{\nu}}-{\boldsymbol \phi}_{{\boldsymbol{\theta}}_1', \boldsymbol{\nu}'}\right\Vert }{\left\Vert {\boldsymbol{\theta}}-{\boldsymbol{\theta}}'\right\Vert +\left\Vert \boldsymbol{\nu}-\boldsymbol{\nu}'\right\Vert }\right\}<\infty.
$$
\end{itemize}

We have (C1) follows from Assumption \ref{as2}(i). (C2)  follows from Theorem \ref{thm4.1}. (C3) and (C4) follow from Assumption \ref{as4}(i)(ii). Thus, we prove \eqref{eqA.18}.

To prove \eqref{eqA.19}, noting that 
$$\widehat{\mathbb{P}}_T{\boldsymbol{\phi}}_{\boldsymbol{\theta},\widehat{\boldsymbol{G}}}-\mathbb{P}_T\boldsymbol{\phi}_{\boldsymbol{\theta},\widehat{\boldsymbol{G}}}=\frac{1}{T}\sum_{t=1}^T\left(\boldsymbol{\phi}_{\boldsymbol{\theta},\widehat{\boldsymbol{G}}}(\widehat{\boldsymbol{F}}_t,\cdots,\widehat{\boldsymbol{F}}_{t+p})-\boldsymbol{\phi}_{\boldsymbol{\theta},\widehat{\boldsymbol{G}}}({\bf R}_0{\boldsymbol{F}}_t,\cdots,{\bf R}_0{\boldsymbol{F}}_{t+p})\right),$$
by Assumption \ref{as4}(iii) and Proposition \ref{prop1}, for any $\epsilon>0$, we have
\begin{eqnarray}
&&\operatorname{P}\left(\sup_{{\boldsymbol \theta}\in {\boldsymbol \Theta}} \left\Vert\widehat{\mathbb{P}}_T{\boldsymbol{\phi}}_{\boldsymbol{\theta},\widehat{\boldsymbol{G}}}-\mathbb{P}_T\boldsymbol{\phi}_{\boldsymbol{\theta},\widehat{\boldsymbol{G}}}\right\Vert<\epsilon\right)\nonumber\\
&\leq&\operatorname{P}\left(\sup_{{\boldsymbol \theta}\in {\boldsymbol \Theta}} \left\Vert\widehat{\mathbb{P}}_T{\boldsymbol{\phi}}_{\boldsymbol{\theta},\widehat{\boldsymbol{G}}}-\mathbb{P}_T\boldsymbol{\phi}_{\boldsymbol{\theta},\widehat{\boldsymbol{G}}}\right\Vert<\epsilon,\left\Vert\widehat{\boldsymbol{G}} - \boldsymbol{G}\right\Vert  \leq \delta\right)+\operatorname{P}\left(\left\Vert\widehat{\boldsymbol{G}} - \boldsymbol{G}\right\Vert  >\delta\right)\nonumber\\
&\leq&\operatorname{P}\left(\sup_{{\boldsymbol{\nu}}\in \mathcal{F}_\delta} \Vert\nabla_{\boldsymbol{x}^\dag}{\boldsymbol \phi}_{\boldsymbol{\theta},\boldsymbol{\nu}}\Vert \max_{1\leq t\leq T}\left\Vert\widehat{\bf{F}}_t-{\bf R}_0{\bf{F}}_t\right\Vert<\epsilon\right)\rightarrow 1.\label{eqA.22}
\end{eqnarray}
Thus, we prove \eqref{eqA.19}.

Lastly, note that
\begin{equation}
    \max_{1\leq t\leq T}\sup_{{\boldsymbol \theta}\in {\boldsymbol \Theta}} |\widehat g_{jt}(x;{\boldsymbol \theta}_H)-\widehat g_{j}(x;{\boldsymbol \theta}_H)| \leq \frac{W(0)}{Tb}.
\end{equation}
\eqref{eqA.26} is proved in Lemma \ref{lemmaB.6}. Thus, we complete the proof of Theorem \ref{thm4.1}. \hfill$\Box$

\section{Technical lemmas}\label{appendixB}
\renewcommand{\theequation}{B.\arabic{equation}}
\setcounter{equation}{0}
\setcounter{lemma}{0}
\renewcommand{\thelemma}{B.\arabic{lemma}}

For simplicity, we write ${\boldsymbol H}_j(\boldsymbol{\theta})$ as ${\boldsymbol h}(\boldsymbol{\theta})$ or ${\boldsymbol h}$ for $j=1,\cdots, K$, and assume that ${\boldsymbol h}(\boldsymbol{\theta})$ is continuous in $\boldsymbol{\theta}$, with $\Vert{\boldsymbol h}(\boldsymbol{\theta})\Vert=1$. Let $${\cal F}=\left\{{\boldsymbol 1}({\boldsymbol h}^\top  \cdot\leq x):\Vert {\boldsymbol h} \Vert=1, {\boldsymbol h}\in \mathbb{R}^K, x\in \mathbb{R}\right\}$$
be a set of indicator functions of half spaces in  $\mathbb{R}^K$ and ${\cal S}^{K-1}=\{{\boldsymbol h}:\Vert {\boldsymbol h} \Vert=1\}$ be the unit sphere in $\operatorname{R}^K$. Define $\mathbb{G}_T(f)=\sqrt{T}({\mathbb{P}}_Tf-{\mathbb{P}}f)$ for $f\in {\cal F}$. We have the following lemma.

\begin{lemma}\label{lemmaB.1}
Suppose that Assumption \ref{as2}(i) is satisfied. For $j=1,\cdots, K$, we have the 
\begin{equation}\label{eqB.1}
\lim_{r\rightarrow0}\lim_{T\rightarrow\infty}\sup_{\operatorname{E}|f'-f''|<r}\mathbb{G}_T(f'-f'')=0, 
\end{equation}
and  
\begin{equation}\label{eqB.2}
\mathbb{G}_T(f)\rightarrow_w \mathbb{G}(f),
\end{equation}
where $\mathbb{G}(f)$ is a zero mean Gaussian process indexed by $f\in {\cal F}$. 

\end{lemma}
\noindent{\bf Proof of Lemma \ref{lemmaB.1}}.
Note that ${\cal F}$ is a uniformly bounded permissible class \citep[see][pp. 21]{S06} with finite Vapnik–\v{C}hervonenkis dimension \citep[see][pp. 833--834]{SW09}, 
and that the mixing coefficients decay exponentially fast. Then, \eqref{eqB.1} is proved by Lemma 2.1 in \cite{AY94} and \eqref{eqB.2} is a direct consequence of Corollary 2.1  in \cite{AY94}. 
\hfill$\Box$
\medskip

 Define $\boldsymbol{\Xi}_R=\sqrt{T}({\bf R}_{NT}-{\bf R}_0)$. By Proposition \ref{prop1}, we have 
\begin{eqnarray}
z_{\boldsymbol{\theta},t}={\boldsymbol h}^\top(
\widehat{\boldsymbol F}_t-{\bf R}_0{\boldsymbol F}_t)&=&{\boldsymbol h}^\top({\bf R}_{NT}-{\bf R}_0){\boldsymbol F}_t+{\boldsymbol h}^\top(\widehat{\boldsymbol F}_t-{\bf R}_{NT}{\boldsymbol F}_t)\nonumber\\
&=&T^{-1/2}\boldsymbol{\Xi}_R{\boldsymbol F}_t + o_P(T^{-1/2}).\nonumber
\end{eqnarray}
 Let $\boldsymbol{\xi}_R$ be a real deterministic matrix corresponding to $\boldsymbol{\Xi}_R$. We restrict $({\boldsymbol \theta}_H,\boldsymbol{\xi}_R)$ in ${\cal B}$, where
\begin{eqnarray*}
    {\cal B}=\left\{({\boldsymbol \theta}_H,\boldsymbol{\xi}_R):{\boldsymbol \theta}_H \in {\boldsymbol \Theta}_H,\Vert\boldsymbol{\xi}_R\Vert\leq \Delta\right\}
\end{eqnarray*}
for some constant $\Delta$. Define 
\begin{equation*}
 z_{t}= z_t({\boldsymbol \theta}_H,\boldsymbol{\xi}_R, \delta_z)= \frac{1}{\sqrt{T}}{\boldsymbol h}^\top\boldsymbol{\xi}_R{\boldsymbol F}_t+\frac{1}{\sqrt{T}}(1+\Vert{\boldsymbol F}_t\Vert)\delta_z
\end{equation*}
for some constant $\delta_z$ and 
$$a_t(x;{\boldsymbol \theta}_H,\boldsymbol{\xi}_R, \delta_z)={\boldsymbol 1}{({\boldsymbol H}_j^\top {\bf R}_0{\boldsymbol F}_t\leq x+ z_t)}-G_{j}(x+ z_t;{\boldsymbol \theta}_H)-{\boldsymbol 1}{({\boldsymbol H}_j^\top {\bf R}_0{\boldsymbol F}_t\leq x)}+ G_{j}(x;{\boldsymbol \theta}_H).$$
When there is no ambiguity, we denote $a_t(x;{\boldsymbol \theta}_H,\boldsymbol{\xi}_R, \delta_z)$ simply as $a_t(x)$ or $a_t$. 
Note that $z_t$ is an $\beta$-mixing process, and so is $a_t$.

\medskip
  Let $x_r$'s be equally spaced grid points ${\cal X}^\epsilon=\{x_r: r=\cdots,-1,0,1,\cdots\}$ satisfying $x_{r}=r \epsilon /\sqrt{T}$. Because ${\boldsymbol \Theta}_H$ is a compact set, we can define ${\boldsymbol \Theta}_H^\epsilon$ as a finite set such that for any ${\boldsymbol \theta}_H$, there exists ${\boldsymbol \theta}_H^\diamond\in {\boldsymbol \Theta}_H^\epsilon$ satisfying $|h({\boldsymbol \theta}_H^\diamond)-h({\boldsymbol \theta}_H)|<\epsilon$.
\begin{lemma}\label{lemmaB.2}
 Suppose that Assumption \ref{as2} is satisfied and $T^{4/\delta_{\varepsilon}+1} \ll N \ll \exp\{T^{1/5}\}$ with $\delta_{\varepsilon}$ defined in Assumption \ref{as2}(v).  We have
$$\sup_{{\boldsymbol \theta}_H\in{\boldsymbol \Theta}_H^\epsilon}\sup_{x\in{\cal X}^\epsilon}\left|\frac{1}{\sqrt{T}}\sum_{t=1}^Ta_t(x;{\boldsymbol \theta}_H,\boldsymbol{\xi}_R, \delta_z)\right|=o_P(1).$$
\end{lemma}

\noindent{\bf Proof of Lemma \ref{lemmaB.2}}. \ \ 
By the moment inequality of the $\alpha-$mixing process \citep[e.g.,][Lemma 1.2.5]{LL97}, and noting that $\operatorname{E}[|a_t|^{2+\gamma}]^{\frac{4}{2+\gamma}}\leq \operatorname{E}[|a_t|^2]$, we have 
\begin{eqnarray}
\operatorname{E}\left[\left(\sum_{t=1}^Ta_t\right)^4\right]
&\leq& 4!\sum_{i=1}^{T-1}\sum_{t=1}^{T-i}\sum_{j\leq i}\sum_{k\leq i}\left|\operatorname{E}\left[a_ta_{t+i}a_{t+i+j}a_{t+i+j}\right]\right|+\nonumber\\
 && 4!\sum_{i=1}^{T-1}\sum_{t=1}^{T-i}\sum_{j\leq i}\sum_{k\leq i}\left|\operatorname{E}\left[a_{t-i-j-k}a_{t-i-j}a_{t-i}a_{t}\right]\right|+\nonumber\\
  && 4! \sum_{i=1}^{T-1}\sum_{t=1}^{T-i}\sum_{j\leq i}\sum_{k\leq i}\left|\operatorname{E}\left[a_{t-k}a_{t}a_{t+i}a_{t+i+j}\right]\right|\nonumber\\
  &\leq& C\left(\sum_{i=0}^{T-1}\sum_{t=1}^{T}i^2(\beta(i))^{\frac{\gamma}{2+\gamma}}\operatorname{E}[|a_t|^{2+\gamma}]^{\frac{4}{2+\gamma}}+
\left(\sum_{i=0}^{T-1}\sum_{t=1}^{T}(\beta(i))^{\frac{\gamma}{2+\gamma}}\operatorname{E}[|a_t|^{2+\gamma}]^{\frac{2}{2+\gamma}}\right)^2\right)\nonumber\\
&\leq&C\sum_{t=1}^{T}\operatorname{E}[|a_t|^2]+CT\sum_{t=1}^{T}(\operatorname{E}[|a_t|^2])^2.\label{eqB.3}
\end{eqnarray}
Following the same argument of the proof of Lemma 1 in \cite{KWXXY19},
we have 
\begin{equation}\label{eqB.4}
\sum_r\sum_{t=1}^T \operatorname{E}[a_t(x_r)^2] \leq 2T+\frac{2T^{1/2}}{\epsilon}\sum_{t=1}^T\operatorname{E}[| z_t|]=O(T),
\end{equation}
and
\begin{equation}\label{eqB.5}
\sum_r\sum_{t=1}^T (\operatorname{E}[a_t(x_r)^2])^2 \leq 2c \sum_{t=1}^T\operatorname{E}[| z_t|]+\frac{2cT^{1/2}}{\epsilon}\sum_{t=1}^T\operatorname{E}[ z_t^2]=O(T^{1/2}),
\end{equation}
where $c=\sup_{{\boldsymbol \theta}_H}\sup_x g_{jt}(x;{\boldsymbol \theta}_H).$
By the Markov inequality and \eqref{eqB.3}--\eqref{eqB.5}, as $T,N\rightarrow\infty,$
\begin{eqnarray}
 \operatorname{P}\left(\sup_{{\boldsymbol \theta}_H\in{\boldsymbol \Theta}_H^\epsilon}\sup_{x \in {\cal X}^\epsilon }\frac{1}{\sqrt{T}}\sum_{t=1}^Ta_t(x;{\boldsymbol \theta}_H)\geq \delta\right)
&\leq&|{\boldsymbol \Theta}_H^\epsilon|\cdot\sum_{r}\operatorname{P}\left(\frac{1}{\sqrt{T}}\sum_{t=1}^Ta_t(x)\geq \delta\right)\nonumber\\
    &\leq&\frac{|{\boldsymbol \Theta}_H^\epsilon|}{\delta^4T^2}\sum_{r}\operatorname{E}\left[\left(\sum_{t=1}^Ta_t(x)\right)^4\right]\nonumber\\
  &=&O_P(T^{-1/2})=o_P(1).\nonumber
\end{eqnarray}
Thus we complete the proof of Lemma \ref{lemmaB.2}. 
\hfill$\Box$
\medskip

\begin{lemma}\label{lemmaB.3}
Suppose that the assumptions in Lemma \ref{lemmaB.2} hold. We have
\begin{equation*}
\sup_{x \in \mathbb{R}}\left|\frac{1}{\sqrt{T}}\sum_{t=1}^Ta_t(x;{\boldsymbol \theta}_H)\right|\leq\sup_{x \in {\cal X}^\epsilon }\left|\frac{1}{\sqrt{T}}\sum_{t=1}^Ta_t(x;{\boldsymbol \theta}_H)\right|+o_P(1),
\end{equation*}
where $o_P(1)$ is uniform in $(\boldsymbol{\theta}_H,\boldsymbol{\xi}_R)\in{\cal B}$.
\end{lemma}

\noindent{\bf Proof of Lemma \ref{lemmaB.3}}.\ , w By the definition of ${\boldsymbol \Theta}_H^\epsilon$ and ${\cal X}^\epsilon$, we have 
\begin{eqnarray}\label{eqB.6}
&&\sup_{{\boldsymbol \theta}_H\in{\boldsymbol \Theta}_H}\sup_{x \in \mathbb{R}}\left|\frac{1}{\sqrt{T}}\sum_{t=1}^Ta_t(x;{\boldsymbol \theta}_H)\right|-\sup_{{\boldsymbol \theta}_H\in{\boldsymbol \Theta}_H^\epsilon}\sup_{x \in {\cal X}^\epsilon }\left|\frac{1}{\sqrt{T}}\sum_{t=1}^Ta_t(x;{\boldsymbol \theta}_H)\right|\nonumber\\
&\leq&\sup_{\Vert{\boldsymbol \theta}'-{\boldsymbol \theta}''\Vert\leq T^{-1/4}}\sup_{|x'-x''|\leq \frac{\epsilon}{\sqrt{T}}} \frac{1}{\sqrt{T}}\sum_{t=1}^T\left[a_t(x';{\boldsymbol \theta}_H'')-a_t(x';{\boldsymbol \theta}_H'')\right]\nonumber\\
&\leq&\sup_{\Vert{\boldsymbol \theta}'-{\boldsymbol \theta}''\Vert\leq T^{-1/4}}\sup_{|x'-x''|\leq \frac{\epsilon}{\sqrt{T}}} R_{1t}(x',x'',\boldsymbol{\theta}_H',\boldsymbol{\theta}_H'')\nonumber\\
&&\sup_{\Vert{\boldsymbol \theta}'-{\boldsymbol \theta}''\Vert\leq T^{-1/4}}\sup_{|x'-x''|\leq \frac{\epsilon}{\sqrt{T}}} R_{2t}(x',x'',\boldsymbol{\theta}_H',\boldsymbol{\theta}_H'')+\nonumber\\
&&\sup_{\Vert{\boldsymbol \theta}'-{\boldsymbol \theta}''\Vert\leq T^{-1/4}}\sup_{|x'-x''|\leq \frac{\epsilon}{\sqrt{T}}} R_{3t}(x',x'',\boldsymbol{\theta}_H',\boldsymbol{\theta}_H''),
\end{eqnarray}
where 
\begin{align*}
& R_{1t}(x',x'',\boldsymbol{\theta}_H',\boldsymbol{\theta}_H'')=\frac{1}{\sqrt{T}}\sum_{t=1}^T\left[G_{j}(x'+z_t(\boldsymbol{\theta}_H');{\boldsymbol \theta}_H')-G_{j}(x''+z_t(\boldsymbol{\theta}_H'');{\boldsymbol \theta}_H'')\right],\\
&R_{2t}(x',x'',\boldsymbol{\theta}_H',\boldsymbol{\theta}_H'')=\frac{1}{\sqrt{T}}\sum_{t=1}^T\left[{\boldsymbol 1}{({\boldsymbol h}(\boldsymbol{\theta}_H')^\top {\bf R}_0{\boldsymbol F}_t\leq x'+z_t(\boldsymbol{\theta}_H'))}-{\boldsymbol 1}{({\boldsymbol h}(\boldsymbol{\theta}_H'')^\top {\bf R}_0{\boldsymbol F}_t\leq x''+z_t(\boldsymbol{\theta}_H''))}\right],\\
&R_{3t}(x',x'',\boldsymbol{\theta}_H',\boldsymbol{\theta}_H'')=\frac{1}{\sqrt{T}}\sum_{t=1}^T\left[{\boldsymbol 1}{({\boldsymbol h}(\boldsymbol{\theta}_H')^\top {\bf R}_0{\boldsymbol F}_t\leq x')}-G_{j}(x';{\boldsymbol \theta}_H')-{\boldsymbol 1}{({\boldsymbol h}(\boldsymbol{\theta}_H'')^\top {\bf R}_0{\boldsymbol F}_t\leq x'')}+ G_{j}(x'';{\boldsymbol \theta}_H'')\right].\label{eqB.7}
\end{align*}
\begin{eqnarray}
R_t(\boldsymbol{\theta}_H)
&=&\sup_{{\boldsymbol \theta}_H\in{\boldsymbol \Theta}_H}\max_r\max_{x\in[x_r,x_{r+1}]}\frac{1}{\sqrt{T}}\sum_{t=1}^T\left[G_{j}(x_{r+1}+z_t;{\boldsymbol \theta}_H)-G_{j}(x+z_t;{\boldsymbol \theta}_H)\right]+\nonumber\\
&&\sup_{{\boldsymbol \theta}_H\in{\boldsymbol \Theta}_H}\sup_{|x'-x''|\leq \frac{\epsilon}{\sqrt{T}}}\frac{1}{\sqrt{T}}\sum_{t=1}^T\left[{\boldsymbol 1}{({\boldsymbol H}_j^\top {\bf R}_0{\boldsymbol F}_t\leq x')}-G_{j}(x';{\boldsymbol \theta}_H)-{\boldsymbol 1}{({\boldsymbol H}_j^\top {\bf R}_0{\boldsymbol F}_t\leq x'')}+ G_{j}(x'';{\boldsymbol \theta}_H)\right]\nonumber\\
&:=&{\boldsymbol\Pi}_{R,1}+{\boldsymbol\Pi}_{R,2}.\label{eqB.7}
\end{eqnarray}
and a reverse inequality holds with $x_{r+1}$ replaced by $x_r$,
For ${\boldsymbol\Pi}_{R,1}$, we have
\begin{equation}\label{eqB.8}
{\boldsymbol\Pi}_{R,1}= O_P(\epsilon\sup_{{\boldsymbol \theta}_H\in{\boldsymbol \Theta}_H}\sup_{x\in \mathbb{R}} g_{jt}(x;{\boldsymbol \theta}_H))=O_P(\epsilon), 
\end{equation}
uniformly in $\boldsymbol{\theta}_H\in \boldsymbol{\Theta}_H$ and therefore in $(\boldsymbol{\theta}_H,\boldsymbol{\xi}_R)\in{\cal B}$. For $|{\boldsymbol\Pi}_{R,2}|$, noting that 
\begin{eqnarray}\label{eqB.9}
&&\left|\operatorname{E}[{\boldsymbol 1}{({\boldsymbol h}({\boldsymbol \theta}_H')^\top {\bf R}_0{\boldsymbol F}_t\leq x')}]-\operatorname{E}[{\boldsymbol 1}{({\boldsymbol h}({\boldsymbol \theta}_H'')^\top {\bf R}_0{\boldsymbol F}_t\leq x'')}]\right|\nonumber\\
&=&|G(x';{\boldsymbol \theta}_H')-G(x'';{\boldsymbol \theta}_H'')|\nonumber\\
&\leq&\left(\sup_{{\boldsymbol \theta}_H\in{\boldsymbol \Theta}_H}\sup_{x\in \mathbb{R}} g(x;{\boldsymbol \theta}_H)+\sup_{{\boldsymbol \theta}_H\in{\boldsymbol \Theta}_H}\sup_{x\in \mathbb{R}}\nabla_{{\boldsymbol \theta}_H} G(x;{\boldsymbol \theta}_H)\right)\left(|x'-x''|^2+\Vert{\boldsymbol \theta}_H'-{\boldsymbol \theta}_H''\Vert^2\right)^{1/2},\nonumber\\
&\leq&  \frac{\epsilon}{\sqrt{T}}
\end{eqnarray}
by Lemma \ref{lemmaB.1} we have $|{\boldsymbol\Pi}_{R,2}|=o_P(1)$.

Therefore, 
$R_t(\boldsymbol{\theta}_H)=o_P(1)$ uniformly in $\boldsymbol{\theta}_H\in \boldsymbol{\Theta}_H$ and therefore in $(\boldsymbol{\theta}_H,\boldsymbol{\xi}_R)\in{\cal B}$.
\hfill$\Box$
\medskip

\begin{lemma}\label{lemmaB.4}
Suppose that the assumptions in Lemma \ref{lemmaB.2} hold. We have
\begin{equation*}
\sup_{{\boldsymbol \theta}_H \in {\boldsymbol \Theta}_H}\sup _{x \in \mathbb{R}}    \left|\frac{1}{\sqrt{T}}\sum_{t=1}^T\left({\boldsymbol 1}({\boldsymbol H}_j^\top \widehat{\boldsymbol F}_t\leq x)-G_{j}(\widehat{\boldsymbol F}_t;{\boldsymbol \theta}_H)-{\boldsymbol 1}({\boldsymbol H}_j^\top {\bf R}_0{\boldsymbol F}_t\leq x)+ G_{j}({\bf R}_0{\boldsymbol F}_t;{\boldsymbol \theta}_H)\right)\right|=o_P(1).
\end{equation*}
\end{lemma}
\noindent{\bf Proof of Lemma \ref{lemmaB.4}}.\ \ Define $\bar{z}_t= {z}_t({\boldsymbol \theta}_H,\boldsymbol{\xi}_R, 0)$ and 
$$\bar{a}_t(x;{\boldsymbol \theta}_H,\boldsymbol{\xi}_R)={\boldsymbol 1}{({\boldsymbol H}_j^\top {\bf R}_0{\boldsymbol F}_t\leq x+ \bar{z}_t)}-G_{j}(x+ \bar{z}_t;{\boldsymbol \theta}_H)-{\boldsymbol 1}{({\boldsymbol H}_j^\top {\bf R}_0{\boldsymbol F}_t\leq x)}+ G_{j}(x;{\boldsymbol \theta}_H).$$
By Proposition \ref{prop1}, we have $\operatorname{P}({\cal B})\rightarrow 1$, as $N,T\rightarrow\infty$. Thus we only need to prove
\begin{equation}\label{eqB.10}
\sup_{{\cal B}}\sup_{x \in \mathbb{R}}\left|\frac{1}{\sqrt{T}}\sum_{t=1}^T\bar{a}_t(x;{\boldsymbol \theta}_H,\boldsymbol{\xi}_R)\right|=o_P(1).
\end{equation}

We divide $(-\Delta, \Delta)$ into $2M$ intervals, $\{C_1,C_2, \cdots, C_m\}$, each with length $\delta:=\Delta/M$. Consequently, for any $({\boldsymbol \theta}_H, \boldsymbol{\xi}_R)$, taking $\boldsymbol{\xi}_R^\diamond$ in the same partition of $\boldsymbol{\xi}_R$, 
we obtains that
\begin{equation*}
\left|z_t({\boldsymbol \theta}_H,\boldsymbol{\xi}_R, \delta_z)-z_t({\boldsymbol \theta}_H^\diamond,\boldsymbol{\xi}_R^\diamond, \delta_z)\right|
\leq(2\delta+\delta^2)\Vert{\boldsymbol F}_t\Vert, 
\end{equation*}
which implies that 
\begin{eqnarray}\label{eqB.11}
z_t({\boldsymbol \theta}_H^\diamond,\boldsymbol{\xi}_R^\diamond, -\delta_z^*)
\leq z_t({\boldsymbol \theta}_H,\boldsymbol{\xi}_R, 0)
\leq 
z_t({\boldsymbol \theta}_H^\diamond,\boldsymbol{\xi}_R^\diamond, \delta_z^*)
\end{eqnarray}
with $\delta_z^*=2\delta+\delta^2$. For simplicity, we denote \eqref{eqB.11} by $z_t^-\leq \bar{z}_t\leq z_t^+$ and let $z_t^\pm$ mean either $z_t^-$ or $z_t^+$. By the monotonicity of the
indicator function, we obtain that
\begin{equation}\label{eqB.12}
\frac{1}{\sqrt{T}}\sum_{t=1}^T\bar{a}_t(x)\leq \frac{1}{\sqrt{T}}\sum_{t=1}^Ta_t^+(x)+T^{-1/2}\sum_{t=1}^T \left(G_j(x+z_t^+;{\boldsymbol \theta}_H)-G_j(x+\bar{z}_t;{\boldsymbol \theta}_H)\right),
\end{equation}
and 
\begin{equation}\label{eqB.13}
\frac{1}{\sqrt{T}}\sum_{t=1}^T\bar{a}_t(x)\geq \frac{1}{\sqrt{T}}\sum_{t=1}^Ta_t^-(x)+T^{-1/2}\sum_{t=1}^T \left(G_j(x+z_t^-;{\boldsymbol \theta}_H)-G_j(x+\bar{z}_t;{\boldsymbol \theta}_H)\right).
\end{equation}
By the mean value theorem, we have 
\begin{eqnarray}
&&T^{-1/2}\sum_{t=1}^T \left|G_j(x+z_t^\pm;{\boldsymbol \theta}_H)-G_j(x+\bar{z}_t;{\boldsymbol \theta}_H)\right|\nonumber\\&\leq & T^{-1/2}\sup_x|g_{jt}(x)|\cdot\sum_{t=1}^T |z_t^\pm-\bar{z}_t|\nonumber\\
&=&O_P(\delta\cdot T^{-1/2}\sum_{t=1}^T \Vert{\boldsymbol F}_t\Vert)=O_P(\delta),
\end{eqnarray}
uniformly for all $x$ and $({\boldsymbol \theta}_H,\boldsymbol{\xi}_R)$.
By Lemmas \ref{lemmaB.2} and \ref{lemmaB.3}, we prove that 
\begin{equation}\label{eqB.15}
\left|\frac{1}{\sqrt{T}}\sum_{t=1}^T\bar{a}_t^\pm(x)\right| = o_P(1),
\end{equation}
uniformly for all $x$ and $({\boldsymbol \theta}_H,\boldsymbol{\xi}_R)$.
Combining \eqref{eqB.12}--\eqref{eqB.15}, we prove \eqref{eqB.10}, and thus we complete the proof.
\hfill$\Box$

\medskip

\begin{lemma}\label{lemmaB.6}
Suppose that the assumptions in Lemma \ref{lemmaB.2} hold. (i) For $j=1,\cdots, K$, we have 
\begin{equation}
\sup_{{\boldsymbol \theta}_H \in {\boldsymbol \Theta}_H}\left|\frac{1}{T}\sum_{t=1}^T\log( g_{j}({\boldsymbol H}_j{\bf R}_0{\boldsymbol F}_t;{\boldsymbol \theta}_H)- {\operatorname{E}[log(g_{j}({\boldsymbol H}_j{\bf R}_0{\boldsymbol F}_t;{\boldsymbol \theta}_H))]}\right|=o_P(1),
\end{equation} 
and 
\begin{equation}
\sup_{{\boldsymbol \theta}_H \in {\boldsymbol \Theta}_H}\left|\frac{1}{T}\sum_{t=1}^T\log( g_{j}({\boldsymbol H}_j\widehat{\boldsymbol F}_t;{\boldsymbol \theta}_H)- \frac{1}{T}\sum_{t=1}^T[log(g_{j}({\boldsymbol H}_j{\bf R}_0{\boldsymbol F}_t;{\boldsymbol \theta}_H))]\right|=o_P(1).
\end{equation} 
(ii) Furthermore, if Assumptions \ref{as3} and \ref{as4}, for $j=1,\cdots, K$, we have
\begin{equation}\label{eqB.18}
\sup_{{\boldsymbol \theta}_H \in {\boldsymbol \Theta}_H}\left|\frac{1}{T}\sum_{t=1}^T\log(\widetilde g_{j}({\boldsymbol H}_j\widehat{\boldsymbol F}_t;{\boldsymbol \theta}_H)- \frac{1}{T}\sum_{t=1}^T[log(g_{j}({\boldsymbol H}_j\widehat{\boldsymbol F}_t;{\boldsymbol \theta}_H))]\right|=o_P(1).
\end{equation} 

\end{lemma}

\noindent{\bf Proof of Lemma \ref{lemmaB.6}}.\ \
(i) The first argument is a direct consequence of ULLN for $\alpha$-mixing process \citep[see][Theorem 2.8]{BW85}. The second argument follows by Assumption \ref{as1}(vii) and Proposition \ref{prop0}.

(ii) By Theorem 6 of \cite{Masry96}, we can prove that 
\begin{equation}\label{eqB.19}
\sup_{{\boldsymbol \theta}_H \in {\boldsymbol \Theta}_H}\sup_x\left|\widehat g_{j}(x;{\boldsymbol \theta}_H)- g_{j}(x;{\boldsymbol \theta}_H)\right|=O_P\left((\frac{\log T}{Tb})^{1/2}+b^2\right)=o_P(1),
\end{equation}
where $\widehat g_j$ is the probabilistic density function of ${\boldsymbol H}_j{\bf R}_0{\boldsymbol F}_t$. The similar result can be find in \cite{HHW25}.

Using the Lipschitz continuity of the kernel \( W \), we obtain,
for any \( x \in \mathbb{R} \) and \( \boldsymbol{\theta}_H \in \Theta_H \),
\begin{eqnarray}
\left| \widetilde g_{j}(x; \boldsymbol{\theta}_H) - \widehat g_{j}(x; \boldsymbol{\theta}_H) \right| 
&\leq& \frac{1}{(T-1)b_j}  \left| \sum_{s=1 }^T\left[W\left( \frac{x - {\boldsymbol H}_j^\top \widehat{\boldsymbol F}_s}{b_j} \right) - W\left( \frac{x - {\boldsymbol H}_j^\top {\bf R}_0 {\boldsymbol F}_s}{b_j} \right)\right] \right|\nonumber\\
&\leq& \frac{\Vert W' \Vert}{b_j^2} \max_{1\leq s \leq T}\left|{\boldsymbol H}_j^\top (\widehat{\boldsymbol F}_s - {\bf R}_0 {\boldsymbol F}_s)\right|.\nonumber
\end{eqnarray}
Thus
\begin{equation}\label{eqB.20}
\sup_{{\boldsymbol \theta}_H \in \Theta_H} \sup_{x \in \mathbb{R}} \left| \widetilde g_{j}(x; {\boldsymbol \theta}_H) -\widehat g_{j}(x; {\boldsymbol \theta}_H) \right| = o_P(1).
\end{equation}
Combining \eqref{eqB.19} and \eqref{eqB.20}, we
\begin{equation*}
\sup_{{\boldsymbol \theta}_H \in \Theta_H} \sup_{x \in \mathbb{R}} \left| \widehat g_{j}(x; {\boldsymbol \theta}_H) - g_{j}(x; {\boldsymbol \theta}_H) \right| = o_P(1),
\end{equation*}
and therefore
\[
 | \widehat g_{j}({\boldsymbol H}_j^\top \widehat{\boldsymbol F}_s; {\boldsymbol \theta}_H) - g_{j}({\boldsymbol H}_j^\top \widehat{\boldsymbol F}_s; {\boldsymbol \theta}_H) | = o_P(1).
\]
By the continuous mapping theorem and Assumption \ref{as2}(vii), we have 
\[
 |\log(\widehat g_{j}({\boldsymbol H}_j^\top \widehat{\boldsymbol F}_s; {\boldsymbol \theta}_H)) - \log(g_{j}({\boldsymbol H}_j^\top \widehat{\boldsymbol F}_s; {\boldsymbol \theta}_H)) | = o_P(1).
\]

Let ${\bf g}({\bf u})$ denote the joint density function of the vector ${\bf R}_0 {\boldsymbol F}_t$, $g_j(u_j)$ be the marginal density of the projection ${\boldsymbol H}_j^\top {\boldsymbol F}_t$, and ${\bf g}_j^\perp({\bf u}^\perp \mid u_j)$ be the conditional density of the orthogonal component ${\bf u}^\perp$ given ${\boldsymbol H}_j^\top {\bf u} = u_j$. Here, ${\bf u}^\perp := ({\boldsymbol H}_j)^\perp {\bf u} \in \mathbb{R}^{K-1}$ denotes the projection of ${\bf u}$ onto the orthogonal complement of the direction ${\boldsymbol H}_j$ with $({\boldsymbol H}_j)^\perp$ being an orthonormal basis of the subspace orthogonal to ${\boldsymbol H}_j$, satisfying $({\boldsymbol H}_j^\perp)^\top {\boldsymbol H}_j ={\bf 0}$ and $({\boldsymbol H}_j^\perp)^\top {\boldsymbol H}_j^\perp = {\bf I}_{K-1}$. Under this representation, the joint density admits the decomposition ${\bf g}({\bf u})=g_j({\boldsymbol H}_j^\top {\bf u}){\bf g}_j^\perp({\bf u}^\perp \mid {\boldsymbol H}_j^\top {\bf u})$. We can show that 
\begin{eqnarray}
\nabla_{{\boldsymbol H}_j}\widetilde g_{j}({\boldsymbol H}_j^\top({\boldsymbol \theta}_H) \widehat{\boldsymbol F}_t;{\boldsymbol \theta}_H)
&=& \frac{1}{Tb_j} \sum_{s=1}^T W'\left(\frac{{\boldsymbol H}_j^\top({\bf R}_0 {\boldsymbol F}_s - \widehat{\boldsymbol F}_t)}{b_j}\right) ({\boldsymbol F}_s - \widehat{\boldsymbol F}_t) \nonumber\\
&=& \frac{1}{b_j} \int W'\left(\frac{{\boldsymbol H}_j^\top({\bf u} - \widehat{\boldsymbol F}_t)}{b_j}\right) ({\bf u} - \widehat{\boldsymbol F}_t) \, {\bf g}({\bf u}) \, d{\bf u} + o_P(1) \nonumber\\
&=& \int_{\mathbb{R}} W'(z) \left( \int_{{\boldsymbol H}_j^\top {\bf u} = z b_j + {\boldsymbol H}_j^\top \widehat{\boldsymbol F}_t} ({\bf u} - \widehat{\boldsymbol F}_t) \cdot g_{|j}({\bf u}_{\perp} \mid z b_j + {\boldsymbol H}_j^\top \widehat{\boldsymbol F}_t) \, d{\bf u}_{\perp} \right) \nonumber\\
&& \quad \cdot g_j(z b_j + {\boldsymbol H}_j^\top \widehat{\boldsymbol F}_t) \, dz + o_P(1),\label{eqB.21}
\end{eqnarray}
which is uniformly bounded over ${\boldsymbol \theta}_H$.
Note that $\max_{1\leq t\leq T}\Vert{\mathbf F}_t \Vert=O_P\left((\log T)^{1/2}\right)$ and that 
\begin{eqnarray}
&&\Vert \nabla_{{\boldsymbol H}_j}\widehat g_{j}({\boldsymbol H}_j^\top({\boldsymbol \theta}_H) \widehat{\boldsymbol F}_t;{\boldsymbol \theta}_H) -  \nabla_{{\boldsymbol H}_j}\widetilde g_{j}({\boldsymbol H}_j^\top({\boldsymbol \theta}_H) \widehat{\boldsymbol F}_t;{\boldsymbol \theta}_H) \Vert\nonumber\\
&\leq&\frac{1}{Tb_j} \sum_{s=1}^T \left\Vert W'\left(\frac{{\boldsymbol H}_j^\top(\widehat{\boldsymbol F}_s - \widehat{\boldsymbol F}_t)}{b_j}\right) (\widehat {\boldsymbol F}_s - \widehat{\boldsymbol F}_t)- W'\left(\frac{{\boldsymbol H}_j^\top({\bf R}_0{\boldsymbol F}_s - \widehat{\boldsymbol F}_t)}{b_j}\right) ({\bf R}_0 {\boldsymbol F}_s - \widehat{\boldsymbol F}_t)\right\Vert\nonumber\\
&\leq& \frac{\Vert W''\Vert}{Tb_j^2} \cdot \sum_{s=1}^T\left\Vert{\boldsymbol H}_j^\top (\widehat{\boldsymbol F}_s - {\bf R}_0 {\boldsymbol F}_s)\widehat{\boldsymbol F}_s\right\Vert +
\frac{\Vert W''\Vert}{Tb_j^2} \cdot \sum_{s=1}^T|{\boldsymbol H}_j^\top (\widehat{\boldsymbol F}_s - {\bf R}_0 {\boldsymbol F}_s)|\left\Vert\widehat{\boldsymbol F}_t\right\Vert\nonumber\\
&&+ \frac{\Vert W'\Vert}{Tb_j^2} \sum_{s=1}^T\left|{\boldsymbol H}_j^\top (\widehat{\boldsymbol F}_s - {\bf R}_0 {\boldsymbol F}_s)\right|\left\Vert\widehat{\boldsymbol F}_t\right\Vert \nonumber\\
&=&O_P\left(\frac{\log T}{Tb_j^2}\left(\frac{1}{T^{1/2}}+\frac{T^{2/\delta_{\varepsilon}}}{N^{1/2}}\right)\right)=o_p(1), \label{eqB.22}
\end{eqnarray}
uniformly over ${\boldsymbol \theta}_H$. Combining \eqref{eqB.21} and \eqref{eqB.22}, we have 
$$\sup_{{\boldsymbol \theta}_H \in \Theta_H}\Vert\nabla_{{\boldsymbol H}_j}\widehat g_{j}({\boldsymbol H}_j^\top({\boldsymbol \theta}_H) \widehat{\boldsymbol F}_t;{\boldsymbol \theta}_H)\Vert=O_P(1).$$

Since both \( \widehat g_{j}({\boldsymbol H}_j^\top({\boldsymbol \theta}_H) \widehat{\boldsymbol F}_s; {\boldsymbol \theta}_H) \) and \( g_{j}({\boldsymbol H}_j^\top({\boldsymbol \theta}_H) \widehat{\boldsymbol F}_s; {\boldsymbol \theta}_H) \) have bounded derivatives with respect to ${\boldsymbol H}_j$ (equivalently, with respect to \( {\boldsymbol \theta}_H \)), they are stochastically equicontinuous. Moreover, since they are uniformly bounded away from zero, and the logarithm function is uniformly continuous on any close interval bounded away from zero, the log-transformed versions are also stochastically equicontinuous. Therefore, by Theorem 22.9 of \cite{D21}, we have
\[
\sup_{{\boldsymbol \theta}_H \in \Theta_H} \left| \log(\widehat g_{j}({\boldsymbol H}_j^\top \widehat{\boldsymbol F}_s; {\boldsymbol \theta}_H)) - \log(g_{j}({\boldsymbol H}_j^\top \widehat{\boldsymbol F}_s; {\boldsymbol \theta}_H)) \right| = o_P(1),
\]
which implies \eqref{eqB.18}. Thus we complete the proof.
\hfill$\Box$

\section{Parameterizing the rotation matrix}\label{appendixC}
\renewcommand{\theequation}{C.\arabic{equation}}
\renewcommand{\theproposition}{C.\arabic{proposition}}
\setcounter{proposition}{0}
\setcounter{equation}{0}
\setcounter{lemma}{0}
\renewcommand{\thelemma}{C.\arabic{lemma}}

Notably, for any nonzero constants $c_1,\cdots,c_K$, the rotation matrix ${\bf H}^{\diamond}=(c_1{\boldsymbol H}_1,\cdots,c_K{\boldsymbol H}_K)$, yields the same value of the likelihood function, as 
\begin{equation}
\widehat L_{T,{\bf b}}(\widehat{\bf F}{\bf H};{\boldsymbol \theta}_H,{\boldsymbol \theta}_C)
=\widehat L_{T,{\bf b}^*}(\widehat{\bf F}{\bf H}^{\diamond};{\boldsymbol \theta}_H^{\diamond},{\boldsymbol \theta}_C),
\end{equation}
with bandwidths ${\boldsymbol b}^*=(|c_1| b_1,\cdots,|c_K| b_K)$. This invariance to scaling indicates a need to normalize or reparameterize ${\bf H}$ to ensure identifiability of the model. To guide the parameterization process, we first outline key properties of the S-vine copula model in the following proposition.
\medskip

\begin{proposition}\label{prop0}
Suppose $({\boldsymbol F}_1,\cdots, {\boldsymbol F}_T)$ follows an S-vine copula model $({\cal V}^F,{\cal C}({\cal V}^F))$. \\
\noindent (i) Letting ${\bf D}$ be any positive definite diagonal matrix, then  $({\bf D}{\boldsymbol F}_1,\cdots, {\bf D}{\boldsymbol F}_T)$ follows the same S-vine copula model $({\cal V}^{F},{\cal C}({\cal V}^{F}))$. \\
\noindent (ii)  Letting ${\bf P}$ be any permutation matrix,  $({\bf P}{\boldsymbol F}_1,\cdots, {\bf P}{\boldsymbol F}_T)$ follows an S-vine copula model $({\cal V}^{PF},{\cal C}({\cal V}^{PF}))$,
with  ${\cal V}^{PF}$ is the vine structure obtained by permuting ${\cal V}^F$.

\noindent (iii)  Letting ${\bf S}_i$ be a diagonal matrix whose $(i,i)-th$ entry is -1 and $(j,j)$-th entry is 1 for $j\neq i$, then  $({\bf S}_i{\boldsymbol F}_1,\cdots, {\bf S}_i{\boldsymbol F}_T)$ follows an S-vine copula model $({\cal V}^{F},{\cal C}^*({\cal V}^{F}))$,
 with ${\cal C}^*({\cal V}^{F})) = \{c_e^*,e\in E_1^F\}\cup \{c_e : e \in E_k^F, k = 2, \cdots, Kp - 1\}$ and $c_e^*=c_e$ if $i \notin e$ and $c_e^*$ is a marginal reflected copula of $c_e$ if $i \in e$.

\end{proposition}
\medskip

Thus, to simplify the estimation process and ensure that the model is identifiable, we can impose a normalization constraint on ${\bf H}$, such as setting one element of each column of ${\bf H}$ to 1, or normalizing each column to have a unit norm.  Alternatively, we can parameterize ${\bf H}$ as
 \begin{equation}\label{eq.rep}
 {\bf H}({\boldsymbol \theta}_H)=\begin{bmatrix}
\sin(\theta_{11}) & \cos(\theta_{11})\sin(\theta_{12})&\cdots&\cos(\theta_{11})\cos(\theta_{12})\cdots\cos(\theta_{1,K-1})\\
\sin(\theta_{21}) & \cos(\theta_{21})\sin(\theta_{22})&\cdots&\cos(\theta_{21})\cos(\theta_{22})\cdots\cos(\theta_{2,K-1})\\
\vdots&\vdots&\vdots&\\
\sin(\theta_{K1}) & \cos(\theta_{K1})\sin(\theta_{K2})&\cdots&\cos(\theta_{K1})\cos(\theta_{K2})\cdots\cos(\theta_{K,K-1})\\
\end{bmatrix}^\top,
\end{equation}
where ${\boldsymbol \theta}_H=({\boldsymbol \theta}_{H,1}^\top,\cdots,{\boldsymbol \theta}_{H,K}^\top)^\top$ and ${\boldsymbol \theta}_{H,i}=(\theta_{i1},\cdots,\theta_{i,K-1})^\top$ for $i=1,\cdots,K$. The parameter vectors ${\boldsymbol \theta}_{H,i}$ for $i = 1, \dots, K$ are arranged in lexicographical order, with $\theta_{i1} \in [0, \pi]$ and $\theta_{ij} \in [0, 2\pi)$ for $j \neq 1$ and $i = 1, \dots, K$. 

The parameterization in \eqref{eq.rep} ensures that each column of ${\bf H}$ has a unit norm, addressing the scaling invariance described in Proposition \ref{prop0}(i). This avoids issues of non-identifiability in the likelihood function due to arbitrary scaling. Additionally, 
the lexicographical order of ${\boldsymbol \theta}_{H,i}$ ensures that the likelihood function remains invariant under permutations, as required by Proposition \ref{prop0}(ii). 
Finally, the constraint $\theta_{i1} \in [0, \pi]$ guarantees that the first element of each column of $\mathbf{H}$ is non-negative, thereby preventing sign changes that could compromise model identifiability, as discussed in Proposition \ref{prop0}(iii).

\begin{remark} \label{rem1}
The parameterization of ${\bf H}$ is not unique \citep[see][for an example]{HHW25} and we do not impose any specific parameterization method. However, a poorly chosen parameterization may lead to underspecification, which, in turn, can result in multiple maxima in the objective function during estimation. This issue is illustrated and discussed further in Section \ref{sec5}.
\end{remark}

%
\end{document}